\documentclass[a4paper,11pt]{article}
\usepackage{tablefootnote}
\usepackage[english]{babel}
\usepackage[T1]{fontenc}
\usepackage[flushmargin]{footmisc}
\usepackage{setspace}
\usepackage{float}
\usepackage{amsmath}
\usepackage{amsfonts}
\usepackage{amssymb}
\usepackage{ae}
\usepackage{placeins}
\usepackage[comma]{natbib}
\usepackage{caption}
\usepackage{graphicx}
\usepackage{adjustbox}
\usepackage{subcaption}
\usepackage{changepage}
\usepackage{tabularx,booktabs}
\usepackage[utf8]{inputenc}

\usepackage[a4paper,left=2.5cm,right=2.5cm,top=3cm,bottom=3cm]{geometry}
\usepackage[colorinlistoftodos]{todonotes}
\usepackage[colorlinks=true, allcolors=blue]{hyperref}
\begin{document} \doublespacing \pagestyle{plain}

	\begin{center}

		{\LARGE The Impact of the \#MeToo Movement on Language at Court \\ A text-based causal inference approach }
		
		{\large 
			\vspace{1.3cm}}
		
		{\large Henrika Langen}\medskip
		
		{\normalsize University of Fribourg, Dept.\ of Economics \bigskip }
	\end{center}
	
	\bigskip

	\noindent \textbf{Abstract:}{\small 
	\ This study assesses the effect of the \#MeToo movement on the language used in judicial opinions on sexual violence related cases from 51 U.S. state and federal appellate courts. The study introduces various indicators to quantify the extent to which actors in courtrooms employ language that implicitly shifts responsibility away from the perpetrator and onto the victim. One indicator measures how frequently the victim is mentioned as the grammatical subject, as research in the field of psychology suggests that victims are assigned more blame the more often they are referred to as the grammatical subject. The other two indices designed to gauge the level of victim-blaming capture the sentiment of and the context in sentences referencing the victim and/or perpetrator. Additionally, judicial opinions are transformed into bag-of-words and tf-idf vectors to facilitate the examination of the evolution of language over time. The causal effect of the \#MeToo movement is estimated by means of a Difference-in-Differences approach comparing the development of the language in opinions on sexual offenses and other crimes against persons as well as a Panel Event Study approach. The results do not clearly identify a \#MeToo-movement-induced change in the language in court but suggest that the movement may have accelerated the evolution of court language slightly, causing the effect to materialize with a significant time lag. Additionally, the study considers potential effect heterogeneity with respect to the judge’s gender and political affiliation. The study combines causal inference with text quantification methods that are commonly used for classification as well as with indicators that rely on sentiment analysis, word embedding models and grammatical tagging.
}

{\small \smallskip }



\bigskip

{\thispagestyle{empty}\pagebreak  }

\section{Introduction}\label{intro_metoo}
The present study evaluates how the \#MeToo movement affected the way sexual offenses are handled within the U.S. justice system by analyzing the language used in judicial opinions published in U.S. state and federal appellate courts between 2015 and 2020. It examines the movement's impact by means of a Difference-in-Differences (DiD) and a panel data-based event study approach. In the DiD approach, the development of language in opinions on sexual offenses is compared to that in opinions on other crimes against persons. Meanwhile, the panel event study approach assesses how the development of language in judicial opinions published by different judges changed as a result of the \#Metoo movement.

This study develops novel text indicators aimed at measuring the extent to which actors in courtrooms employ language that implicitly shifts responsibility away from the perpetrator and onto the victim. The primary indicator measures the frequency with which the victim is mentioned as the grammatical subject, drawing upon psychology research which suggests that victims tend to be assigned greater blame when they are often mentioned as the grammatical subject. (see e.g. \cite{Stricklandetal2014} and \cite{Niemietal2016}) Additionally, the study develops two other indicators to capture the development of victim-blaming language, which build on the word2vec algorithm and the SentiWordNet 3.0 lexicon (\cite{JatowtDuh2014}) respectively, as well as an approach to use text vectorization methods typically used for text categorization in order to quantify evolution of language over time. The study then demonstrates how the different text quantifiers can be integrated into a DiD and an event study approach in order to estimate the causal effect of the \#MeToo movement on language in judicial opinions.   In doing so, this study contributes to the recently growing literature on text-based causal inference. By applying this approach to analyze the impact of the \#MeToo movement on language in courts, the study addresses a research  question  that has not been previously examined.

The study proceeds as follows. Section \ref{Background_metoo}  reviews the current state of literature on text analysis, particularly the emerging field of text-based causal inference, and provides background information on the \#MeToo movement and the U.S. justice system. Section \ref{textdata_metoo} describes the corpus of judicial opinions  examined in this study. The following section, Section \ref{textquantification_metoo}, illustrates how the judicial opinions are quantified by means of various indicators and text vectorization approaches. Section \ref{identification_metoo} outlines the identification strategy underlying this paper and discusses the assumptions necessary to identify the impact of the \#MeToo movement using a DiD model and a panel data-based event study approach. The subsequent section (\ref{Results_metoo}) summarizes the results and Section \ref{conclusion_metoo} concludes.


\section{Background}\label{Background_metoo}
\subsection{Text-Based Causal Inference}
While there are several socio-economic and legal studies on text classification based on Natural Language Processing (NLP) and also some NLP-based analyses on language development, literature on text-based causal inference is scarce. The social science literature on text classification ranges from studies on differences in the linguistic style between posts in different online communities (\cite{KhalidSrinivasan2020}) and between comments on \#MeToo articles in different news outlets (\cite{Rhoetal2018}) to studies that develop classifiers for political speeches in order to predict the speaker's ideology (\cite{Yuetal2008}) or identify his/her sentiment towards the topic discussed in the speech (\cite{AbercrombieBatista2018}). In the legal domain, \cite*{Hausladenetal2020} developed a document classifier for  judicial opinions from U.S. circuit courts that classifies the opinions according to the predicted ideological direction (conservative vs. liberal) of the decision. In addition, there are several studies on classifying legal documents by topic (see, e.g., \cite{Undaviaetal2018}, \cite{Filtzetal2019} and \cite{Alekseevetal2019}). 

The evolution of language over time has been analyzed using quantifiers for the context in which words are used (see, e.g., \cite{Kulkarnietal2015}, \cite{Hamiltonetal2016} and \cite{FrermannLapata2016}) as well as based on indicators for the sentiment of that context (see, e.g., \cite{JatowtDuh2014} and \cite{Hellrichetal2018}). \cite*{NguyenRose2011} analyze how new members in a medical forum gradually adapt their language to the forum's linguistic standards during their first year of forum participation. They quantify the language in posts by using indicators that measure lexical features of posts, including counts of colloquial words and words falling within various psychological, topical and linguistic categories as identified by the Linguistic Inquiry and Word Count (LIWC) tool. 


Literature on text-based causal inference has emerged only in recent years. Some studies integrate NLP elements into causal inference for confounding adjustment (see \cite{Keithetal2020} for a review). \cite*{Robertsetal2020} for example develop a framework for estimating treatment effects which combines text-based matching and confounding adjustment based on text. They estimate how a scholar's gender affects the number of citations of his/her publications, while controlling for and matching based on the content of publications. Other studies that rely on text-based confounding adjustment include those by \cite*{Sallin2021} and \cite*{Veitchetal2020}. \cite*{Mozeretal2020} and \cite*{Fieldetal2020} propose a text matching approach based on distance metrics rather than text classification. \cite*{Woodetal2018} integrate text classifiers into causal inference in order to tackle problems with missing data and measurement error. 

Other studies use text as treatment or outcome. \cite*{Ornaghietal2019} analyze how a judge's score on an indicator of gender-stereotyped language affects their decisions on women’s rights’ issues. \cite*{Tanetal2014} assess how wording in tweets affects the number of re-tweets. To do so, they apply different text vectorization approaches  and quantify wording by means of different indicators of tweet features such as sentiment, lexical distinctiveness and readability, in order to identify words that in-/decrease re-tweet propensity.  Similarly, \cite*{Deshpandeetal2022}, \cite*{Pryzantetal2020}, \cite*{Wang2019}, \cite*{FongGrimmer2016} and \cite*{Feuerriegeletal2015} evaluate how to estimate causal effects of wording, semantics and lexical choices in texts. 

\cite*{Egamietal2018} developed a sample splitting framework for estimating treatment effects with text as outcome, building on the classification of outcome texts based on a model trained in the training sample. \cite*{Sobolev2018} assesses how troll activity that promotes a pro-government agenda on Russian social media affects the evolution of online discussions using a regression discontinuity approach. To do so, he models the development of conversations on social media as changes in the mixture of topics with topics identified through NLP-based classification. Other examples of studies with texts as an outcome include a study by \cite*{Chandrasekharanetal2017} on how Reddit’s 2015 anti-harassment policy affected the usage of hate speech, as well as an analysis by \cite*{Pavalanathanetal2018} on the effect of tagging articles as not written in a ``neutral point of view'' on the development of lexical patterns in the labeled articles. 

Among the studies on text-based causal inference with text as the outcome, there is, to my knowledge, no study yet using a panel event study approach, and so far only one study that applies a DiD approach, namely the study by \cite*{Chandrasekharanetal2017} on the effects of Reddit’s anti-harassment policy on hate speech use. This study compares the development of an indicator that measures the frequency of terms typically associated with hate speech posts between individuals who were members in a group where the anti-harassment policy was violated and those who were not. Hate speech-related terms are identified from conversations in groups that were banned in the context of the introduction of the 2015 anti-harassment policy.

\subsection{The \#MeToo Movement and its societal, cultural and political impact}
After starting as an online campaign against sexual harassment, \#MeToo soon evolved into a movement that led to extensive and sustained media coverage of the prevalence of sexual violence in society. It prompted discussions about abuse of power, rape myths and the importance of supporting victims of sexual assault. 


The phrase ``Me Too” was initially coined by social justice activist Tarana Burke who began using this phrase in 2006 to campaign for the empowerment of sexual violence victims, particularly among women of color.  It gained widespread attention in late 2017 after sexual misconduct allegations against Harvey Weinstein were exposed in a New York Times exposé by \cite*{KantorTwohey2017} published on October 5th, 2017, followed just days later by an investigative article in The New Yorker on the same matter (\cite{Farrow2017}). On October 15th, actress Alyssa Milano tweeted a post encouraging victims of sexual harassment and assault to come forward by using the hashtag \#MeToo in order to raise awareness about the prevalence of sexual violence in society. The \#MeToo hashtag rapidly spread as more and more people around the world shared their experiences. On Twitter, the hashtag was tweeted about 300,000 times on the day after Milano’s post, reached a peak of 750,000 tweets within 24h and was used on average more than 55,000 times per day during the year following the initial tweet (\cite{AndersonToor2018}). On Facebook, the \#MeToo conversation peaked at 4.7 million participating users within 24 hours, who engaged with over 12 million posts, comments, and reactions (\cite{SantiagoCriss2017}). 

The discussion on social media grew into a movement that led to protest marches in the U.S. and around the globe, resulted in extensive and sustained media coverage of the issues of sexual violence and abuse of power, and shaped the public discourse in the months following October 2017. \cite*{Caputietal2019} estimate that in the first 8 months after the movement’s emergence, the number of google searches on sexual harassment and assault exceeded the expected amount by some 86\%. According to the Women’s Media Center, the number of articles on sexual assault and harassment in a sample of 14 leading U.S. newspapers was more than double the pre-\#MeToo average in November 2017 and still exceeded the pre-\#MeToo average by 30\% some 10 months after the onset of the movement (\cite{EnnisWolfe2018}). \cite*{Time2017} named the “Silence Breakers” - victims of sexual harassment or assault who came forward and thereby started the global dialogue on sexual violence - as its 2017 “Person of the Year”, i.e., as the person or group that most influenced the events of the year, according to the magazine.

\begin{figure}[h]
	\captionsetup{font=footnotesize}
	\centering
	\includegraphics[width=0.6\textwidth]{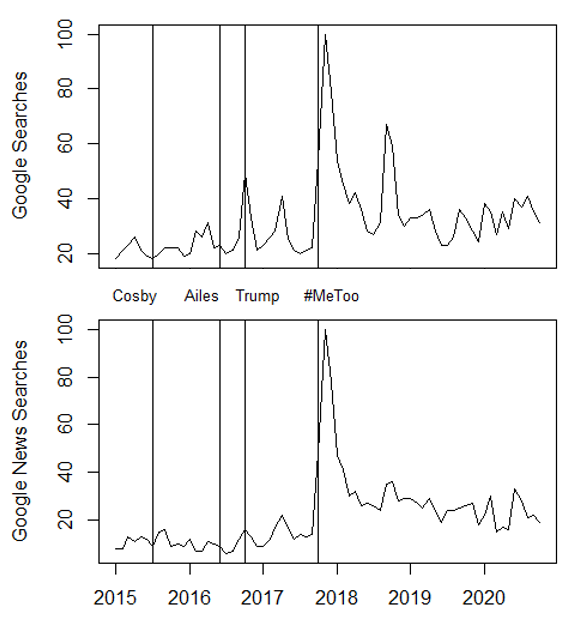}
	\caption[Google searches for ''sexual assault'' \& ``sexual harassment'']{\label{fig:googletrends_metoo} Searches on Google and Google News in relation to the highest point in the period January 2015 to November 2020. The vertical lines indicate when the sexual harassment/assault allegations against stand-up comedian Bill Cosby, former U.S. President Donald Trump and Fox News CEO Roger Ailes became public.}
\end{figure}

About 65\% of social media users report having regularly encountered at least some content related to sexual harassment or assault on social media platforms in the months following the start of the \#MeToo movement, with little difference across demographic groups (\cite{AndersonToor2018}). The observation that a large share of society has been confronted with the issue of sexual violence is also reflected in Google search data. Following \cite*{Caputietal2019} and \cite*{LevyMattsson2020}, I look at how often the terms ''sexual assault'' and ``sexual harassment'' were searched for on Google  in the U.S. during the study period. Figure \ref{fig:googletrends_metoo} shows that public interest in these topics has never been higher than at the start of the \#MeToo movement. Additionally, Figure \ref{fig:googletrends_metoo} also shows how often these terms were searched for on Google News, with a similarly pointed peak of searches in October 2017. The onset of the movement also saw a sharp increase in the number of articles in traditional media covering sexual harassment and assault, as shown both in an analysis of four major U.S. newspapers conducted by \cite*{LevyMattsson2020} and in a study published by the Women's Media Center (\cite{EnnisWolfe2018}). Following \cite*{LevyMattsson2020}, Figure \ref{fig:googletrends_metoo} further highlights that the \#MeToo movement, rather than the Weinstein scandal, generated the heightened interest in issues related to sexual harassment and assault. This is evident from the fact that scandals involving men with similar or even higher profile than Harvey Weinstein did not attract nearly as much public attention as the \#MeToo mmovement.

While bringing changes in attitudes and responses to sexual violence, the \#MeToo movement has also provoked some disillusionment and backlash. On the one hand, several surveys indicate that both men and women have observed others to avoid sexually harassing behavior in response to \#MeToo, with these changes being obvious in surveys focusing on workplace conduct as well as conduct in public or private settings (\cite{Keplingeretal2019}; \cite{JacksonNewall2018}; \cite{Careerarc2020}; \cite{Greenfield2018}). Other surveys, however, suggest that many employers expect or have experienced men to become more cautious in their interactions with female colleagues, potentially hindering women's professional advancement (see, e.g., \cite{Atwateretal2019}; \cite{Frenchetal2021}; \cite{McGregor2019}; \cite{BertottiMaxfield2018}; \cite{NBSWSJ2017}). 

Furthermore, several articles argue that \#MeToo (almost) exclusively benefited affluent white women while other groups, such as women from lower socioeconomic backgrounds and women of color, lack the public outreach, support systems and financial security necessary for generating public attention and dealing with potential backlashes (see, e.g., \cite{FilebornLoneyHowes2019}; \cite{Kagaletal2019}; \cite{Taub2019}). This criticism of demographic disparities in victim outreach  is also supported by several studies (see e.g. \cite*{Muelleretal2021} and \cite{Evans2018}). The under-representation of sexual violence victims from certain demographic groups, however, does not necessarily translate into a lesser impact of the movement on those groups, as e.g. studies by  \cite*{Palmeretal2021} and \cite{LevyMattsson2020} as well as survey data from the Pew Research Center  (\cite{Pew2018}) show. 


The societal, cultural and political impact of the \#MeToo movement has been analyzed extensively with mixed findings about the movement's influence. There are studies showing that the \#MeToo movement did not significantly increase self-reported interest in political participation (\cite{Castleetal2020}), that the portrayal of male entrepreneurs in Swedish media changed only marginally after the onset of the movement (\cite{Jernbergetal2020}), that the \#MeToo movement caused a significant increase in the propensity of female dating platform users in South Korea to decline dating requests (\cite{Yoonetal2020}) and that it induced a change in gender norms in Swedish-language tweets (\cite{Moricz2019}). A study by \cite*{KlarMcCoy2021} reveals that support for the \#MeToo movement is associated with a stronger belief in the sexual misconduct allegations against Donald Trump among Democrats but not among Republicans. \cite*{Aitetal2020} examined how the movement affected women’s perception of safety by comparing the development of men's and women's perceptions of safety in subway stations with a DiD approach, finding a significant decline in women's perceptions of safety after \#MeToo. 

In the study most relevant to the present one, \cite*{LevyMattsson2020} assess the effect of the \#MeToo movement on the propensity to report a sexual offense to the police. They apply a triple-difference approach over time, across 31 OECD countries, and between sexual and non-sexual offenses. For countries where the \#MeToo movement attracted a great deal of attention, they find that it caused a significant 10\% increase in the number of reported sexual offenses during the first six months after the movement began. A DiD analysis comparing the development of reported sexual and non-sexual crimes in the U.S. reveals a similar long-term effect for the 15 months following Alyssa Milano's tweet. The study further shows that the movement's impact on the reporting of sexual offenses was similar across regions and socioeconomic groups in the U.S., but that it did not affect the reporting of all types of sexual assault equally: reports of rape and fondling increased, while reports of statutory rape and sodomy remained unaffected.  Finally, the study assesses the impact of the movement on sexual offense arrests, finding that the increase in reported sex offenses in the U.S did not bring about a similar surge in the number of arrests. Rather, the authors estimate that the movement raised the number of arrests by only about 6\%, which they attribute to the fact that the movement had a stronger effect on cases with a low likelihood of arrest  due to delayed reporting, less severe offenses, or insufficient evidence. The study does not provide any conclusions about the effect of the movement on convictions. 

Despite the numerous studies examining the societal, cultural, and political implications of the \#MeToo movement, there is, to the best of my knowledge, no quantitative study on its influence on the justice system. The present study aims to assess whether and to what extent the movement has brought about a change in the way cases of sexual violence are addressed in court and how the justice systems deals with victims.


\subsection{Victim Blaming}\label{victimblaming_metoo}
Besides raising awareness about the prevalence of sexual harassment and assault in society, the \#MeToo movement has also fueled discussions about the acceptance of rape myths. For this reason, the judicial opinions under examination in this study are assessed specifically with respect to the  reinforcement of rape myths and victim blaming.

The term ``rape myths'' refers to stereotypical and inaccurate beliefs about sexual assaults that are prevalent in society. These myths are frequently used to redirect blame from the perpetrator to the victim and to downplay the seriousness of the incident. Rape myths may not always be explicitly stated; they can also be reinforced through rhetorical techniques or by highlighting particular details when discussing sexual assault. For example, studies show that sexual violence reports that emphasize external circumstances and the victim's behavior can increase the likelihood of readers to accept rape myths and view the victim as partly responsible for the assault (see, e.g., \cite{Bohner2001}, \cite{Franiuketal2008} and \cite{McCoy2004}). 

Victim blaming and reinforcement of rape myths can be observed in a variety of contexts, albeit with varying degrees of explicitness. While often more bluntly expressed in social media and other informal contexts (see, e.g., \cite{Suvarnaetal2020} and \cite{SuvarnaBhalla2020} for studies on identification of victim-blaming language in informal contexts), victim blaming and reinforcement of rape myths are also present in traditional media (see, e.g., \cite{Sacks2018}, \cite{Northcuttetal2019} and \cite{Franiuketal2008}) and within the justice system. In court, there are general victim rights and sexual assault-specific regulations, such as rape shield laws, that to some extent prohibit explicit victim blaming, stigmatization, and stereotyping. Yet, several studies, based on testimonies from various actors, such as victims, barristers, sex crime investigators and independent observers, suggest that victim blaming and rape myths are still prevalent in courts (see, e.g., \cite{Temkinetal2018}, \cite{Spencer2018}, \cite{SmithSkinner2012}, \cite{Temkin2000}). Although it is often the defense that introduces rape myths for strategic purposes, some judges do not always intervene and may even endorse such arguments (\cite{Temkinetal2018}, \cite{Ehrlich2012}). In addition, many judges tend to employ terminology of affection and consensual sex in cases where the perpetrator is familiar to the victim (\cite{Ehrlich2012}).

\subsection{The U.S. Justice System}\label{justicesystem_metoo}

The legal documents examined in this paper are judicial opinions from 51 U.S. state and federal appellate courts. Appellate courts review legal cases that have already been heard in a lower court (trial court) after a party  appeals the trial court's decision. In civil cases, both parties have the right to appeal the trial court's decision, while in most states, only the defendant can appeal in criminal cases. To appeal a decision, the appealing party (the appellant) must file a brief, i.e., a written argument setting forth the facts and arguing why the trial court's decision was erroneous, to which the other party (the appellee) must respond with an appellee's brief.

The appellate court does not usually admit new evidence or witnesses; it may rule solely on the basis of the written briefs or after hearing oral arguments. The appellate court often issues what is called a judicial opinion, i.e., a written decision outlining the court's reasoning; it is usually written by a single judge and reviewed by the other judges on the panel. If a judge disagrees, they can issue a dissenting opinion. Judges who disagree with the reasoning but agree with the result can issue a concurring opinion. Sometimes, judges issue an unsigned opinion called a per curiam opinion (\cite{ABA2019}).

As the U.S. legal system is based on common law, it can constantly - without the intervention of legislators - produce new doctrines and let others fade away.  To achieve this, judges simply adapt to emerging circumstances and interpret the law accordingly, thus establishing precedents (\cite{Harper2016}). In the U.S. judicial system, not all opinions are considered potential precedents. It distinguishes between precedential opinions, which the authoring court deems to have sufficient precedential value to be published so that others may cite them as precedents, and nonprecedential opinions, primarily intended for the involved parties and citable by others only as persuasive rather than binding authority.

\section{Text Data}\label{textdata_metoo}
The judicial opinions used in this study were obtained from CourtListener, an archive of court data operated by \cite*{FreeLawProject2020}. The CourtListener database collects judicial opinions issued by state and federal courts from various sources. The body of opinions for this study is restricted to precedential opinions from state and federal appellate courts. The reason for this choice is that the corpus of precedential opinions available in the CourtListener database is complete for the available courts, unlike that of non-precedential opinions, i.e., by restricting the sample to precedential opinions, no selection problems arise. In addition, precedents reflect developments in courts and case law. As integral components of the body of law, they contribute to the continuous development of the legal system.

\subsection{The Body of Judicial Opinions}\label{judicialopinions_metoo}
The body of judicial opinions examined in this paper includes opinions from 51 courts. Most judicial opinions of state and federal appellate courts have a similar structure. They usually begin with a summary of the trial court's ruling, followed by the arguments of defense and prosecution that were presented and admitted before the trial court, as well as the appellate brief filed by the defense. They typically conclude with reasoning and the decision of the appellate judges. Thus, the opinions reflect the atmosphere during the trial court hearing and the appeal proceedings, as well as the attitude of the defense and the judges towards the victim.

To single out the opinions on sexual offenses and those on other crimes against persons from the full body of precedential opinions, I take advantage of the fact that the opinions have a similar structure and usually begin with an introduction specifying the offense(s) for which the offender was convicted in trial court. The opinions can therefore be reliably classified based on term search in the introduction. To distinguish opinions on sexual-violence related cases and those on other cases of interpersonal crimes, I rely on regular-expression-based (\texttt{regex}-based) search of the legal terms for such crimes\footnote{The legal terms for sexual offenses differ strongly across  U.S. states (see https://apps.rainn.org/policy/\#report-generator for state-specific terms and definitions); therefore, different state-specific sets of legal terms are used. The legal terms for  offenses of interpersonal violence are more homogeneous across the states; consequently, the same set of terms is used for all courts where the legal terms are obtained from the following site: https://www.criminaldefenselawyer.com/topics/crimes-against-persons)} in the introduction of the respective opinion. 

As the headers of the opinions differ between courts, their divisions and even the judges who author the opinions, the introductions cannot be reliably identified based on \texttt{regex} rules. Therefore, the first third of each opinion but no more than 5,000 characters are defined as the introduction. The introductions of all opinions available at CourtListener are searched for matches to legal terms related to sexual violence; the remaining opinions are then searched for matches to other interpersonal-violence related legal terms. While certainly not adequate to accurately classifying opinions into crime categories, the identification procedure described above ensures that opinions are selected into the sample based on the same rules throughout the observation period. Focusing on the introduction of opinions prevents erroneously classifying opinions to one of the two groups when an opinion mentions crimes from a party's past.

Using the \texttt{regex} procedure described above, I can identify 43,088 opinions on crimes against persons published between January 2015 and November 2020, including 15,307 on sexual offenses and 27,781 on non-sexual offenses against persons. The number of opinions per court is provided in Table \ref{tab:descCourt_metoo}.

\section{Text Quantification}\label{textquantification_metoo}
In order to assess the impact of the \#MeToo movement on language in court, the judicial opinions need to be quantified. The quantifiers described in Section \ref{victimblamingindicators_metoo} aim at quantifying the amount of victim blaming in each opinion, while the text vectorization methods outlined in Section \ref{textvectorization_metoo} are later used to capture the general evolution of language in judicial opinions.

\subsection{Victim-Blaming Indicators}\label{victimblamingindicators_metoo}
To quantify the extent of victim blaming in judicial opinions, three indicators are constructed based solely on sentences in which the victim or perpetrator is named. These indicators aim at capturing the extent to which opinions contain wording that implicitly shifts some blame from the perpetrator onto the victim, where such wording may come from the defense, the prosecution, or the judges involved in the case. The first indicator represents a novel approach based on findings from the field of psychology and will be the primary focus here. The remaining two indices assess the context in which the victim and perpetrator are discussed, where the first one measures the sentiment of the context, while the second one more comprehensively assesses the linguistic environment in which they are referenced.

In the appellate court opinions, the victim is often referred to by his/her name or by the term ``victim''. The appealing party, i.e., the person found guilty in trial court,  is usually called the ``appellant'' or by his/her name, but may also be referred to as the ``petitioner''. As the victims' and the appellants' names are not clearly identifiable, only sentences containing the words ``victim'', ``appellant'' and ``petitioner'' as well as inflected forms of these words are considered in the construction of the indicators.

\noindent
\textbf{The Semantic Role of Victim: Subject vs. Object}

\noindent
The first victim-blaming indicator captures the semantic structure of the sentences in which the victim is mentioned.  This approach is grounded in psychological research, which suggests that grammatical structure can convey an author's perception of a fact and influence how recipients perceive it. \cite*{Niemietal2016} asked participants in an experiment on Amazon's Mechanical Turk (MTurk) to read fictional reports of rape in which either the victim or the perpetrator was the grammatical subject in some 75\% of the sentences. Participants who read reports in which the victim was primarily the grammatical subject were more likely to shift some responsibility for the assault to the victim. These findings are in line with a study by \cite*{Stricklandetal2014} in which study participants were asked to judge the intentionality of the grammatical object and subject in a set of sentences that were ambiguous in terms of intentionality. Study participants attributed significantly more intentionality to the grammatical subject than to the object.

The above findings also apply to sentences and texts written in passive voice.  In an experiment by \cite*{Bohner2001}, study participants were asked to describe an uncommented video showing a rape scene and to complete a questionnaire measuring rape myth acceptance. The study reveals that describing the scene primarily in the passive voice (with the victim as the  subject) is positively correlated with attributing responsibility to the victim. \cite*{Henleyetal1995} confronted study participants with fabricated news reports on violence against women written in either active or passive voice. They found that males but not females rated the perpetrator's responsibility higher after reading reports in the active voice.  

These findings suggest developing a victim-blaming indicator that captures the semantic role of the victim in judicial opinions. I construct an indicator that measures the relative frequency of sentences in which the victim functions as grammatical subject, where the semantic roles of all words are identified using the Python package \texttt{spacy}. 

\noindent
\textbf{Sentiment Orientation}

\noindent
The second indicator aims to capture the sentiment orientation of sentences mentioning the perpetrator. \cite*{JatowtDuh2014} developed such an indicator based on SentiWordNet, a database with information on word sentiments\footnote{The English SentiWordNet 3.0 contains more than 100,000 words, each of which is assigned sentiment scores for positivity and negativity. Since many words have different meanings/senses depending on the context in which they are used, the SentiWordNet dataset contains a separate entry for each meaning of a word. The different word meanings are ranked according to how frequently the word is used with the different meanings.}. \cite*{JatowtDuh2014} propose to calculate sentiment scores for a word of interest as the average of the positivity or negativity scores of all context words (the words surrounding the word of interest). As the meaning of the context words is not identifiable without deeper content analyzes, they take the weighted average of the scores for all meanings of each context word, with the weights calculated based on the meaning ranks from the SentiWordNet dataset. 

To assess the sentiment orientation of sentences in which the perpetrator is named, I determine the negativity score as suggested by \cite*{JatowtDuh2014}. I consider the words that are at most five words away from the mention of the perpetrator and occur in the same sentence. The reason for focusing on the context words surrounding the mention of the perpetrator is that positively connotated context words of the victim may not point at the absence of victim-blaming language, as, e.g., comments on the victim's clothing or attitude towards the perpetrator are common examples of rape myth reinforcement.

\noindent
\textbf{Word Embedding}
\noindent

In a third approach to identifying the effect of the \#MeToo movement on the use of victim-blaming language, the development of the context in which the words ``victim'' and ``appellant''/``petitioner'' appear is assessed by means of Word2Vec\footnote{Word2Vec is a method for vectorizing words in such a way that each word vector (of predefined length) captures the context in which the represented word usually appears, and thereby its semantic and syntactic properties. There are two approaches to learning the vector representation for each word in a text corpus. In the Common Bag of Words (CBOW) approach, the Word2Vec neural net takes context words (the words surrounding the unknown target word in a sentence) as input and returns probabilities for each word in the model vocabulary to appear in the given context. In the Skip-Gram approach, the neural network takes single words as inputs and predicts their context. In both approaches, the inputs are passed through a hidden layer that is constantly updated in order to optimize the returned prediction probabilities. Once the model is trained, the context vectors can be extracted from the hidden layer (\cite{Mikolovetal2013}).}, a neural network-based word embedding method. The Word2Vec algorithm is trained for each calendar quarter and for sexual and personal crimes separately. In order to make the Word2Vec models comparable, they are aligned using Compass Aligned Distributional Embeddings (CADE). As suggested by \cite*{JatowtDuh2014}, I calculate the cosine similarity between the vectors of the first quarter of 2015 (for the words ``victim'' and ``appellant'') and those for each other quarter, both for the group of opinions on sexual offenses and the group of non-sexual offenses, in order to be able to assess the development of the word embedding vectors.  

Since the word embedding approach only identifies one vector per time period and group, this approach only allows for estimating the effect of the \#MeToo movement, but not the uncertainty of the estimate, i.e., the standard errors. 




\subsection{Text vectorization}\label{textvectorization_metoo}
For the analysis of the linguistic development in the judicial opinions, the opinions are vectorized by means of the Bag of Words (BoW) and the term frequency - inverse document frequency (tf-idf) models, both of which are frequently used for text classification. The resulting vectors are then used to quantify the development of the language in court over time, as described further below.

\noindent
\textbf{The Bag of Words Approach}

\noindent
The Bag of Words (BoW) algorithm transforms text documents into fixed-length vectors. Each vector represents the relative frequencies of words in one document, with the length of these vectors determined by the number of unique words in the entire corpus of text documents, known as the model vocabulary. There are a number of words in judicial opinions, such as ``court'', ``trial'' or ``judge'', that appear in almost all opinions and do not provide any insight into the evolution of the language used in court. To place greater emphasis on salient words that may reflect linguistic developments in court opinions, I not only determine the BoW representation of opinions based on the entire vocabulary, but also construct a second set of BoW vectors based only on a vocabulary that includes words that occur in less than 95\% of all cases. This alternate BoW approach will in the following be referred to as reduced-corpus BoW. The tf-idf model described in the next section goes in a similar direction as this reduced-corpus BoW, since both aim at reducing the influence of frequently used words in the vector representations of judicial opinions.

\noindent
\textbf{The Term Frequency - Inverse Document Frequency Approach}

\noindent
The tf-idf model represents text documents as fixed-length vectors just as the BoW model, only that the vector elements of the tf-idf vectors are computed as the relative frequency of each word in a document multiplied by the logarithmically scaled inverse proportion of documents in the corpus that contain this word. This way, the words that are characteristic of one or a set of document(s) are given a higher weight, while vector elements are comparably small if the corresponding word occurs in (almost) every document.

When applied to the corpus of judicial opinions without further text processing, the tf-idf model yields the problem that names of persons, places and organizations that naturally only occur in one or a small set of opinion(s) are given particularly large weights, even though they are not relevant to the documents' content. To circumvent this, I apply the Stanford PoS Tagger\footnote{The Stanford PoS Tagger is a probabilistic conditional log-linear model that - based on lexical features of words as well as the context in which they appear - tags each word in a text as corresponding to a grammatical category such as verb, noun, proper name, etc.} to identify and exclude those words that correspond to names of persons, places and organizations before applying the tf-idf model.

\noindent
\textbf{Assessing the Development of Text Vectors over Time}

\noindent
The high-dimensional vectors determined with the three approaches outlined above are then projected using principal component analysis (PCA) in order to reduce the dimensionality and thereby the computational complexity, while preserving as much of the variance of the original text vectors as possible. Since the dimensionality reduction is applied to all text vectors together, the resulting vectors are suitable for computing similarity between the opinions they represent. \cite*{JatowtDuh2014} take a similar approach to evaluate the evolution of word context vectors over time. 

The data is reduced to as many principal components as needed to capture 90\% of the variation in the original opinion vector dataset. The resulting reduced vectors have 64 (BoW), 65 (reduced corpus BoW) and 36 (tf-idf) dimensions respectively. Studies on text classification usually reduce the data to only 1-5 dimensions to avoid capturing noise. The goal of the present study, however, is not to classify texts, but rather to detect even minor changes in the language used in judicial opinions. These minor changes may indicate, for example, a change in the treatment of victims in court, while in the context of the thematic classification of the comments, they may be considered noise. 

After this data reduction step, I average the vectors of all opinions from the first half of 2015 for each court separately and calculate the distance between these 2015 average vectors and all other opinion vectors of the corresponding court (from opinions published between July 2015 and November 2020) in order to quantify each opinion by its distance from the average of the first half of 2015. To do this, I choose the $L_1$ distance metric (also known as Manhattan distance) as proposed by \cite*{Aggarwaletal2001}. They show that for data of 20 dimensions and more the $L_1$ metric is the best distance measure in terms of contrasting two points. Other distance measures as well as other dimensionality choices will also be considered as part of the robustness checks.

\section{Identification}\label{identification_metoo}

The empirical strategy outlined above is applied to assess the impact of the \#MeToo movement on language in court when dealing with cases of sexual violence.  The outcome, courtroom language, is expected to reflect changes in the treatment of sexual offenses and their victims, as well as in the atmosphere at such trials, that are not due to directly measurable, exogenously imposed reforms, but rather to changes in the attitudes of the parties involved. Language in judicial opinions can shift as the parties involved in a trial change the way they describe sex offenses and how they address victim and offender, where these changes may be both a conscious or unconscious expression of attitudinal shifts. Judges may also change to more conscientiously apply rules of court, codes of conduct and rape shield statutes, with rape shield statutes being laws designed to protect victims of sexual offenses by, for example, prohibiting evidence relating to the victim's past sexual behavior. Moreover, as the U.S. legal system is a common law system, judges may begin to interpret other laws differently, providing the impetus for new doctrines. Such changes in the interpretation of law are most visible in precedential opinions and are likely to be reflected in the language used in those opinions. All of these changes in the way cases of sexual offenses and their victims are handled in court due to changing attitudes are not directly measurable, but rather must be gleaned from written opinions.

By examining various indicators that quantify the extent of victim blaming in court, I place particular attention on how the \#MeToo movement has affected the treatment of victims in court.  Just as with the more general language quantifiers, victim blaming indicators can capture both conscious and unconscious manifestations of attitudinal shifts, with the semantic indicator likely capturing primarily unconscious change, while the other two indicators of victim blaming capture both types of attitudinal shifts. The \#MeToo movement  may be defined as the collective action of sexual violence victims who used the phrase and hashtag ``Me Too'' for publicizing their experiences of sexual harassment and assault in order to point out the prevalence of sexual violence in society.

The effect of the \#MeToo movement on the language at court is assessed by means of a DiD approach and an event study approach both of which are outlined further below. To account for the fact that the process between the initial trial court hearing and the publication of a judicial opinion from an appellate court often takes several months, sometimes years, I particularly look at the effect of \#MeToo on court opinions published at least one year after the movement began; that is, while the estimates for the year following the movement's onset are also reported, a stronger emphasis is put on subsequent years, particularly November 2018 through April 2020, when the Covid-19 pandemic hit the United States. While appeals must be filed promptly after publication of the trial court's decision (usually within 30 days), the process in appellate courts often takes considerably longer: the U.S. Courts of Appeals report the median disposition time, i.e., the time between the filing of an appeal and the appellate court's decision,  to be 8.6 months for 2015.\footnote{see https://www.uscourts.gov/news/2016/12/20/just-facts-us-courts-appeals} Thus, opinions published at least one year after the movement's onset are likely to contain closing arguments, trial court judgements, appellee and appellant briefs, as well as appellate court reasoning and decisions written after the \#MeToo movement began. However, as trial court proceedings can also last for many months or even years, opinions on cases with such lengthy trial court proceedings may also contain citations from or references to early trial court hearings that took place before \#MeToo.  However, the fact that the opinions may include parts of hearings from before \#MeToo should be taken into account when interpreting the estimated effect, as this could lead to an underestimation of the movement's impact on language in the courts.


\subsection{Difference-in-Differences Approach}

I apply a DiD approach both with two and multiple time periods in order to identify the Average Treatment Effect on the Treated (ATET) (see, e.g., \cite{Snow1856}, \cite{CardKrueger1994} and \cite{AcemogluAngrist2001}). The sample consists of all judicial opinions on cases that can be classified as ``crimes against persons''. Opinions classified as treating sexual crimes form the treatment group, i.e., the set of opinions that are affected by the \#MeToo movement, while opinions on other crimes against persons serve as control group.  One advantage of choosing the sample of ``crimes against persons'' is that all crimes against persons involve a victim and a perpetrator, which is important for detecting victim-blaming language. Further, the comparison of sexual offenses and other crimes against persons is also common in the literature on victim blaming and rape-myth acceptance (see, e.g., \cite{Reichetal2021}, \cite{Bieneck2011} and \cite{LevyMattsson2020})).

In order to identify the ATET of the \#MeToo movement using a DiD approach, certain assumptions must hold. For simplicity, I will discuss these assumptions for the simple case of two time periods, but the discussion is easily transferable to the multiple time period approach. 

\textbf{Common Support Assumption}

\noindent
The common support assumption states that for each post-treatment observation from the treatment group, there must be comparable observations in the other three groups, i.e., the pre-treatment observations from the treatment and the control group, as well as the post-treatment observations from the control group. Translated to the present study, there must be opinions in these three groups that are comparable to post-treatment opinions on sexual offenses in terms of court and judge characteristics. This assumption is likely to hold because I only consider judicial opinions from courts that handle both sexual offense cases and cases of other crimes against persons. Further, judges are usually assigned their cases randomly, which is why the characteristics of judges in all 4 groups should be similar. 

\textbf{No Anticipation Assumption}

\noindent
Then, the no anticipation assumption states that the treatment may not have any effect on the outcome in pre-treatment periods, which would have been the case if judges and other actors in court had anticipated the movement and changed their behavior accordingly before the movement went viral. This assumption is also likely to hold as the \#MeToo movement was launched immediately after the sexual assault allegations against Harvey Weinstein came to light, and to an extent that no one anticipated. 

\textbf{Common Trend Assumption}

\noindent
The third assumption is the common trend assumption. It states that in absence of the treatment, outcomes in the treatment and control group would follow a parallel trend, or in other words, that the gap between the outcome in the two groups would be constant over time.  

There are various circumstances that may challenge the validity of this assumption. For one, legislative reforms are likely to affect language in court, as they, for example, bring about changes in how certain crimes are sanctioned, what evidence is admitted in court, or what role parties and witnesses may take in the court process.  To the best of my knowledge, and as noted by \cite*{LevyMattsson2020}, there have been no major legislative changes on crimes against persons (neither on sexual nor on non-sexual offenses) in the United States during the years under study. There are only some states that have enacted laws prohibiting the use of non-disclosure agreements which prevent victims of sexual harassment or assault from speaking out. It is unlikely that this legislative change had a direct effect on the language used in sexual offense lawsuits, as such non-disclosure agreements while designed to prevent sexual harassment charges, have no bearing on what victims say in court when a crime is tried. The prohibition of non-disclosure agreements, however, may have increased the number of sexual offense reports, as victims may have come forward who would have been prevented from speaking out in the absence of these laws. This issue of changes in the number of reported cases of sexual violence is discussed further below in the context of the fourth assumption. 

A second potential violation of the Common Trend Assumption are other exogenous changes in the judges' attitudes toward sexual offense cases, which may result from other sexual harassment or assault scandals, other movements, or even the release of a book or film that addresses sexual violence. All of these events could potentially draw public attention to the issue of sexual violence and lead to a change in attitudes toward sexual violence cases and their victims. These concerns can be mitigated by looking at search history, traditional media and social media data from the United States (see Section \ref{Background_metoo}. All of these data show that at no other time during the period studied was as much attention drawn to sexual harassment and assault as in the weeks following the onset of the \#MeToo movement.

Finally, the language in trials on sexual offenses might generally, even in absence of the\#MeToo movement, evolve faster than that in trials on other crimes against persons. Against this, one can argue that all crimes against persons are relatively ``old'' crimes, i.e., crimes that are unlikely to change in nature and have long been considered crimes, thus necessitating less frequent reinterpretation of corresponding laws (unlike, e.g., cybercrimes). However, recent decades have witnessed a trend toward more gender-sensitive language in parts of society, which is likely to exert a greater impact on discussions about sexual offenses than on those about other crimes against persons. If this trend toward more gender-sensitive language is also perceptible in court, the parallel trend assumption may be violated. Furthermore, various feminist developments in society over recent decades might have encouraged judges to more frequently reinterpret the law on sexual offenses, even if it was not for the \#MeToo movement. Both these issues would lead to a violation of the parallel trend assumption and certainly suggest interpreting the results from the DiD approach with some reservation. However, the parallel trend assumption can and will be tested by means of placebo tests, in order to rule out serious violations of the common trend assumption.


 \textbf{Stable Unit Treatment Value Assumption (SUTVA)}

The SUTVA assumption, finally, rules out spill-over effects between observations. In the present study this implies that the \#MeToo movement must solely influence the language used in opinions related to sexual offenses, without affecting language in opinions concerning other crimes against individuals. Furthermore, the SUTVA rules out compositional changes in the treatment or control group over time. 

Spill-over effects might arise insofar  as the \#MeToo movement might have not only altered how courtroom actors perceive and treat victims of sexual offenses but also potentially led to changes in how victims of other violent crimes are treated. This could result in an underestimation of the movement's impact on the use of victim-blaming language. However, aside from the fact that \#MeToo discourse did not tackle victim blaming in the context of crimes other than sexual offenses, studies have shown that victim blaming is more prevalent in the context of sexual offenses than in that of other crimes (\cite{bieneck2011blaming}), indicating a greater potential for developing a language of compassion toward the victim in sexual offense cases.

Arguing for the validity of the second implication of SUTVA, namely the assumption that there are no compositional changes in either the treatment or control group, presents a greater challenge. The composition of sexual offense cases heard in court may have changed as the \#MeToo movement led to an increase in the reporting of such crimes. \cite*{LevyMattsson2020} show that the \#MeToo movement has brought about an increase in reports of sexual offenses, which in turn induced a slight, albeit statistically significant, rise in arrests related to sexual offenses. Some of these additional arrests (that arguably would not have happened in absence of the \#MeToo movement) may have resulted in convictions, some of which in turn may have been appealed and thus become part of the sample. Similarly, there may be cases in my sample that went to trial only because of the prohibition on non-disclosure agreements that some states enacted in response to \#MeToo. 

Although the increase in cases in itself is not problematic for identifying the effect of \#MeToo on language in court, it may have led to changes in the composition of sex offenses addressed within the judicial opinions in my sample. This, in turn, would be troublesome, as such compositional changes would likely result in language shifts that cannot be attributed to a change in how a given case is treated in court. Concerns about \#MeToo-induced compositional changes in the present sample can  be debunked to at least some extent: For one, the findings by \cite*{LevyMattsson2020} suggest that the set of additional sexual offense reports includes a disproportionate number of comparatively lighter crimes and cases with less pressing evidence. They show that the rise in reports of sexual offenses and subsequent arrests/summonses is mainly driven by an increase in reports of sexual harassment, which rarely lead to criminal trials even when evidence is sufficient. Based on survey data, the authors argue that this is a result of the movement reshaping victims' perception of the severity of the experienced sexual offense. In addition, the authors note that the movement had a particularly strong effect on the reporting of cases that occurred at least one month before being brought to the police, i.e., in cases that are more difficult to prove in court.  In both cases, that is, low severity sexual offenses and cases without pressing evidence, the likelihood of a criminal trial is low, which in turn reduces the chances of these cases reaching an appeals court and ending up in my sample.


A look at Table \ref{tab:composition_metoo} reveals that the share of sexual offense opinions across the sample is constant over time, with 35-36\% of opinions relating to sexual offenses each year. The table also reports the share of sexual offense opinions that address different sexual offenses\footnote{The opinions are categorized based on which sexual offense-related terms could be identified in the opinion's introduction, i.e., one opinion may be counted in more than one category}. To draw on the results by \cite*{LevyMattsson2020} for robustness checks, the terms identified in the introductions are grouped into six categories of  sex offenses similar to those defined by the FBI National Incident-Based Reporting System. In addition to the five categories that match those used by \cite*{LevyMattsson2020}, I also include a subcategory of the broad category of sexual assault, namely sexual assault of children and minors, to examine whether an increase in reports from a particular group may have been offset by a decrease in reports from another group. The table indicates that the composition of sexual offense cases does not show any substantial shifts over time, with none of the sexual offense types having a constantly in- or decreasing share over time, also when particularly looking at the years 2017 to 2019 for which \cite*{LevyMattsson2020} find a significant increase in reports. The table also reports the p-values for the difference proportion of each sexual offense category in 2015 and every other year, controlling for court fixed effects.  The p-values indicate that there are no statistically significant differences between the years under study. When not controlling for court fixed effects, a few of these differences are moderately statistically significant, both before and after the \#MeToo movement, suggesting that the slight inter-temporal differences in the composition are attributable to general variations in the number of opinions per court and differences across courts in the composition of opinions. These results suggest that if there are \#MeToo-induced compositional changes in the composition of sexual offense opinions in the sample, they are small and not statistically significant.

\begin{table}[ht]
	\captionsetup{font=footnotesize}
	\centering
	\adjustbox{width = \textwidth, center}{%
		\begin{tabular}{lcccccc}
			\toprule
			& 2015 & 2016 & 2017 & 2018 & 2019 & 2020\\ 
			\midrule
			Share of Sexual Offense Opinions & 0.361 & 0.356 & 0.353 & 0.353 & 0.357 & 0.352 \\ 
			&  & (0.503) & (0.646) & (0.865) & (0.888) & (0.275) \\ 
			\midrule
			Composition of Sexual Offenses: &  &  &  &  &  &  \\ 
			Sexual Assault & 0.834 & 0.815** & 0.786 & 0.817 & 0.824 & 0.827* \\ 
			&  & (0.932) & (0.015) & (0.644) & (0.773) & (0.098) \\ 
			Sexual Assault on a Minor/Child & 0.028 & 0.028 & 0.028 & 0.032 & 0.026 & 0.036 \\ 
			&  & (0.745) & (0.26) & (0.502) & (0.506) & (0.345) \\ 
			Statutory Sexual Assault & 0.134 & 0.151 & 0.148 & 0.141 & 0.134 & 0.155 \\ 
			&  & (0.982) & (0.616) & (0.977) & (0.42) & (0.731) \\ 
			Sodomy & 0.091 & 0.114 & 0.102 & 0.084 & 0.103 & 0.113 \\ 
			&  & (0.034) & (0.316) & (0.964) & (0.639) & (0.361) \\ 
			Fondling & 0.306 & 0.315 & 0.325 & 0.339 & 0.32 & 0.338 \\ 
			&  & (0.483) & (0.533) & (0.241) & (0.49) & (0.454) \\ 
			Sexual Harassment & 0.002 & 0.006 & 0.007 & 0.006 & 0.004 & 0.006 \\ 
			&  & (0.57) & (0.405) & (0.553) & (0.846) & (0.438) \\ 
			\midrule
			\# Opinions (Total) & 6474 & 7685 & 7111 & 7223 & 7883 & 6713 \\ 
			\# Sexual Offense Opinions & 2333 & 2737 & 2510 & 2548 & 2812 & 2367 \\
			\bottomrule
	\end{tabular}}
	\caption[Composition of sexual-violence related opinions, by crime type \& year]{Shares of sexual-violence related opinions that deal with different crime types (by year), as well as the p-value for the differance between the 2015 share and the share in the respective year (with court fixed effects). Note: For 2016-2020, the year y is defined as November y-1 to October y in order to have a clear cut at the onset of the \#MeToo movement in November 2017; the year 2015 only consists of the months January-October 2015. Significance levels: * p<0.1, ** p<0.05, *** p<0.01.}\label{tab:composition_metoo}
\end{table}

To check for robustness of my estimates against the small changes in the sample composition regarding courts, the DiD analysis is complemented by an Inverse Probability Weighting (IPW) DiD approach (\cite{Abadie2005}), in which I weigh all control observations and the pre-movement sexual offense opinions to have the same distribution of courts as the group of post-treatment sexual offense opinions. \footnote{IPW based on offense categories would be problematic in that the offense categories of the control and treatment groups do not overlap by design.} 

Further, I conduct a second robustness check, building on the results of \cite*{LevyMattsson2020}. The authors find that the \#MeToo movement had a small but statistically significant effect the reporting of some types of criminal sexual offenses, while others, namely sodomy and rape, were not affected. I therefore re-run my analyses, considering only opinions on sexual offenses where the number of reports was not affected by the \#MeToo movement. 

Nevertheless, there may still be \#MeToo-induced shifts in the composition of cases that cannot be controlled for or accounted for in robustness checks: the composition of offenses of a given type could still change in terms of the strength of the evidence and/or the severity of the offense.  This would be the case if the \#MeToo movement had led to an increase in trial court cases involving sexual offenses and, at the same time, a decrease in the proportion of convicted offenders who appeal their convictions. In this case, the proportion of appeal cases with inconclusive evidence may have increased, which could have accelerated the development of language in the sex offense sample and thus biased upward the estimated effects of movement on language development, i.e., the effect estimates from the text vectorization approaches. In contrast, the estimates for the impact of the movement on the use of victim-blaming language would in this scenario constitute lower bounds of the actual decline in the use of such language, since the more reasons there are to doubt the credibility of the victim or the seriousness of the incident, the more likely it is that victim-blaming language will be used.

In the DiD approach, I control for court fixed effects, since the  court- or state-specific laws and rules, as well as the terminology therein, are likely to differ. In addition, I control for the word count to account for the fact that there are some very short and formal opinions in the sample that have little or no flexibility in how they are written. However, I also report the estimates for when no controls are included. The reported standard errors are heterogeneity-robust and clustered at the court level (\cite{Zeileis2004}, \cite{Bertrand2004}).

\subsection{Event Study Approach}

I complement the DiD analysis with an event study approach, in which I assess the development of the text quantifiers before and after the \#MeToo movement in a panel setting. In doing so, I attempt to address the problem that the set of opinions on non-sexual offenses may represent an imperfect control group (see discussion above). In the event study approach, I examine the judge-specific developments in the text quantifiers and victim blaming language indicators before and after the onset of the \#MeToo movement. The purpose of this event study application is not to identify the causal effect of the movement, but rather to observe whether there was a shift in the overall development of language in sex offense cases and the use of victim-blaming language from before to after the movement. 

To apply the event study approach, I take advantage of the fact that for 9 courts, accounting for roughly 30 \% of the opinions in my sample, the names of the judge who wrote the opinion is provided in the CourtListener database. For the remaining sexual offense opinions, I obtain the judge's name based on a semi-automatic approach, i.e., by formulating court-specific \texttt{regex}-rules to identify the authoring judge, which I then check manually based on the context from which they were drawn. In cases where more than one judge is named as the author of an opinion, the name mentioned first is selected as the authoring judge, as is the case in the CourtListener database. Opinions written per curiam, i.e., in the name of the court rather than the judge(s), as well as opinions in which no authoring judge is named are excluded from the sample in the event study setting (they account for less than 1\% of sexual offense opinions). 

The names of the so identified judges are cleaned up, i.e., spellings of the same name and title in opinions of the same court are aligned. Of course, the entire process of identifying the authoring judge has many potential sources of error. For one thing, there could be two judges with the same name in the same court whose opinions will be attributed to one and the same individual. Second, when an opinion is authored by more than one judge, the order in which the judges are named does not necessarily say anything about the writing share of the judges, thus the judge named first and selected by me is not necessarily the primary author. It can be stated, however, that in the vast majority of opinions one single judge is named as the author and opinions with more than one authoring judge are the exception. Finally, both, the judge names identified by CourtListener and those identified by me may occasionally be incorrect. However, since there is no reason to believe that the number of erroneous judge names is time-dependent, this solely affects the estimation by introducing some additional noise, but does not lead to a systematic bias in the estimators.  

I identify 1,382 judges, who, on average, publish roughly 11 sexual opinions during the observation period, with the median number of opinions being 4, i.e., there are a few judges who authored a large amount of opinions (up to 184 opinions) while many others only published a handful opinions during the study period. For each judge in the sample, I calculate the six-month average of each text quantifier and victim blaming indictator\footnote{The 6-month text quantifiers are calculated as the 6-month means of the judge's opinions from the respective court's opinion vectors in H1-2015, expressed relative to the median distance. The 6-month victim blaming indicator is calculated as the sum of mentions of the victim as the subject of a sentence divided by the total number of mentions of the victim in each 6-month period.} to obtain a panel data structure with one or no observation per individual and time period. To avoid losing too many observations, I do not exclude all observations for which data is missing in any time period, but keep all judges who published at least one observation before the onset of the \#MeToo movement and at least one observation a year or more after the start of the movement. Then, I apply a Fixed Effects (FE) approach while weighting the observations by the number of opinions they were calculated from. Through weighting the observations,  I account for the fact that the judges differ greatly in how many sex offense opinions they publish per six-month period, which makes some judges much more important for the development of language in court than others.

Although some information is lost by averaging the observations per judge and 6-month period, the panel approach might eliminate some of the noise typical of text analyses by increasing the amount of text per observation. On the other hand, however, it requires me to exclude several observations from the sample (3843 opinions authored by 785 judges) because I do not have observations from either before or after \#MeToo for the authoring judges. By excluding judges who, for whatever reason, do not frequently publish precedential opinions on sex offenses, important information may be lost. Further, this panel approach captures only part of the evolution of language in court. Changes due to retirement or dismissal, and replacement of judges are obviously not captured in this panel approach.

\subsection{Assessing Effect Heterogeneity}

Finally, I also assess whether there is evidence of effect heterogeneity with respect to a judge's gender or  political affiliation, as well as with regard to the political orientation of the state in which a court is located. This is because different studies on victim blaming and rape myths acceptance show that females are less likely to accept rape myths and shift the blame for an assault upon the victim (e.g., \cite{Pinciotti2021}, \cite{Russell2017}, \cite{Davies2009}, \cite{Schneider2009}). \cite*{Boux2016} finds that this is particularly true for female Democrats. 

Most studies in which participants were confronted with a sexual assault scenario found that men were more likely than women to blame the victim and show signs of rape myth acceptance, while other studies found no significant effect of gender on the likelihood of victim blaming (see \cite{Gravelinetal2019}, \cite{GrubbTurner2012} and \cite{SuarezGadalla2010} for reviews). Other research shows that study participants with politically conservative views are more likely to (partially) blame the victim for a sexual assault  (\cite{Andersonetal1997}, \cite{LambertRaichle2000}). For the judicial context, \cite*{Boux2016} finds that female Democratic judges are less likely to use rape myths than male judges (regardless of political affiliation), while her results show no significant difference between female Republican judges and male judges.

In addition to the differences in victim blaming and  rape myth acceptance noted above, there is also evidence that the perception of and reaction to the \#MeToo movement differ across genders and political camps.  \cite*{Castleetal2020} found in a poll that Democrats were more likely than Republicans to say they were aware of the movement and mobilized by it. An analysis of the members of Congress' communications on their public Facebook pages reveals that in the wake of the \#MeToo movement, far more female than male members addressed the issue of sexual violence in their posts, with this pattern evident in both political parties (\cite{AndersonToor2018}).

In light of these findings, it is interesting to assess whether judges' language in court is affected differently by the \#MeToo movement depending on their gender and political affiliation. The research cited above suggests that female and/or politically liberal judges were more receptive to the \#MeToo movement and more willing to change their behavior. Then again, these judges may have been more cautious in their choice of words prior to the movement and may have already avoided language that implied victim blaming, leaving them little room for change toward language that attributed less blame to the victim.

The analysis of treatment effect heterogeneity by judge characteristics is restricted to courts for which the names of the authoring judge is available on CourtListener. Information on the judges' gender and political affiliation is obtained from ballotpedia, an online encyclopedia on American politics and elections. Ballotpedia only provides the political affiliation of some judges. To determine the political affiliation of the other judges, I draw on the party for which the judge ran in the judicial election or the political affiliation of the politician who appointed the judge, depending on the process used to select judges. For about 21\% of the judges no political affiliation can be determined. To assess effect heterogeneity with respect to a state's political orientation, I categorize those states  as predominantly Democratic (Republican) that were won by the Democratic (Republican) party in at least three of the four 2008-20 presidential elections. Opinions from swing states and courts at the supra-state level are excluded for this analysis.

\section{Results}\label{Results_metoo}
\subsection{Victim Blaming Indicators}

\begin{figure}
	\captionsetup{font=footnotesize}
	\centering
	\begin{subfigure}{.49\textwidth}
		\centering
		\includegraphics[width=\linewidth]{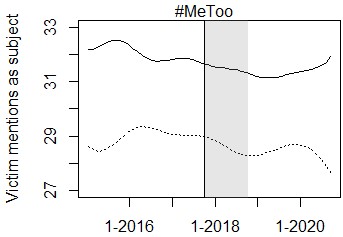}
	\end{subfigure}
	\begin{subfigure}{.49\textwidth}
		\centering
		\includegraphics[width=\linewidth]{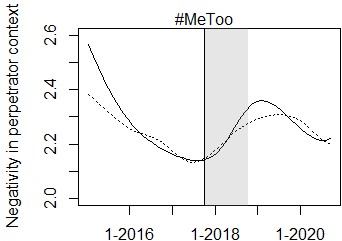}
	\end{subfigure}
	\caption[Development of victim blaming indicators in treatment \& control group]{Kernel-smoothed plot of victim blaming indicators, with the solid line representing sexual offense opinions and the dotted line representing the control group. Curves were smoothed using the default settings of the \texttt{sm.regression} function in R. The semantics indicator measures mentions of the victim as a subject as a percentage of total mentions, and the sentiment indicator measures the negativity score of the context words of mentions of the offender, ranging from 0 to 100, with the words being more negative the higher the score.}
	\label{fig:smoothedVictimBlaming_metoo}
\end{figure}

\begin{figure}
	\captionsetup{font=footnotesize}
	\centering
	\begin{subfigure}{.49\textwidth}
		\centering
		\includegraphics[width=\linewidth]{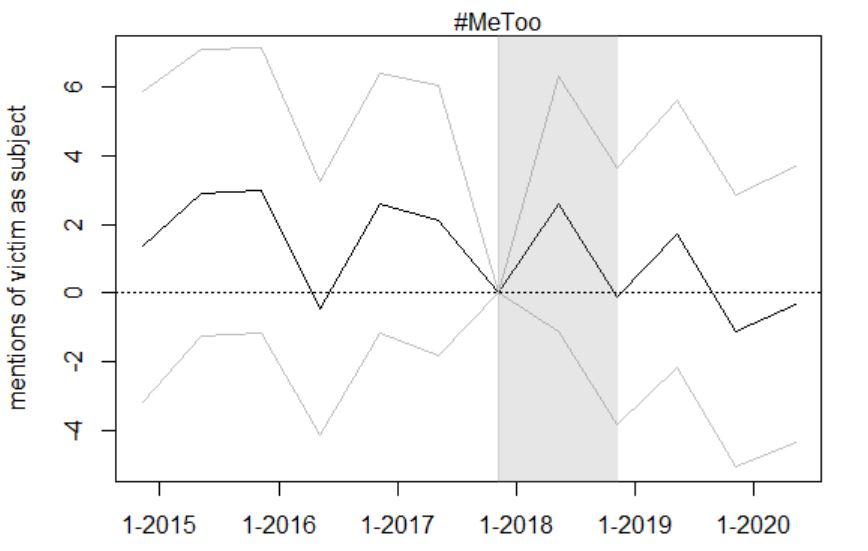}
	\end{subfigure}
	\begin{subfigure}{.49\textwidth}
		\centering
		\includegraphics[width=\linewidth]{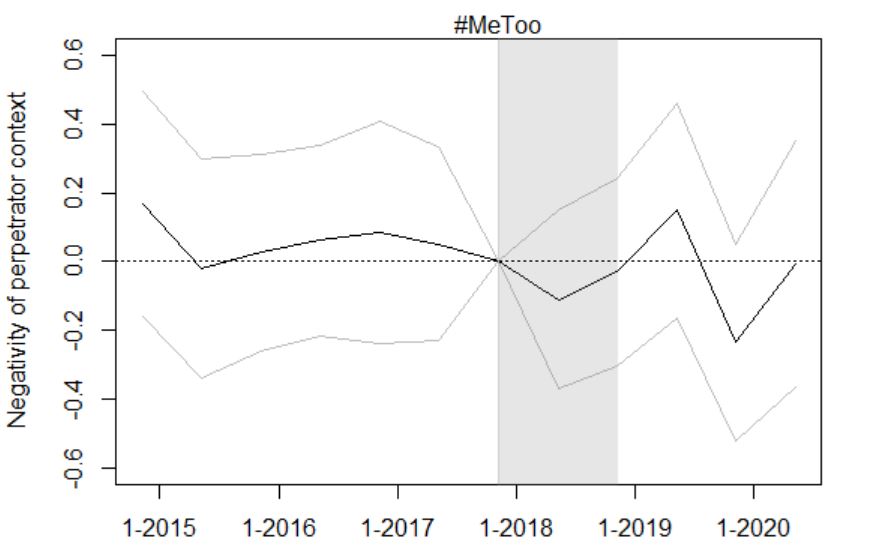}
	\end{subfigure}
	\caption[Event study estimates for evolution of victim blaming indicators]{\label{fig:event_victimblaming_metoo} Event study estimates for the  evolution of the victim blaming indicators in sexual offense opinions. The gray lines indicate the 90 percent confidence interval around the estimates.}
\end{figure}

Figure \ref{fig:smoothedVictimBlaming_metoo} shows the development of the victim blaming semantics and sentiment indicators in opinions on sexual offenses and on other crimes against persons. Neither plot suggests any substantial differences between the development of these two indicators in the treatment and the control group. Further, the DiD estimates provided in Table \ref{tab:victimblaming_metoo} do not indicate a significant impact of the \#MeToo movement on either of the two indicators. The DiD estimates for the semantics indicator suggest a slight, but not statistically significant, \#MeToo-induced decline in the number of victim mentions as a grammatical subject and hence a decline in victim blaming. The estimates for the sentiment indicator are also not statistically significant and even indicate, contrary to the research hypothesis, a decrease in the use of words with negative connotations in the context of mentions of the perpetrator. Likewise, the event study approach (see Figure \ref{fig:event_victimblaming_metoo})  shows no substantial changes in either of the two victim blaming indicators.

\begin{table}
	\captionsetup{font=footnotesize}
	\centering
	\adjustbox{width = 0.85\textwidth, center}{%
		\begin{tabular}{lcccccc}
			\toprule
			& \multicolumn{3}{c}{Share of Victim as Subject} &            \multicolumn{3}{c}{Negativity of Offender Context}       \\ 
			\midrule
			sex cr. x post & -0.687  &  &    & -0.079  &  &  \\ 
			& (1.087) &  &  & (0.056) &  &  \\ 
			sex cr. x '19 &  & -0.446  &    &  & 0.002  &  \\ 
			&  & (1.088) &    &  & (0.064) &  \\ 
			sex cr. x '20 &  & -0.872  &   &  & -0.087  &  \\ 
			&  & (1.340) &    &  & (0.057) &  \\ 
			sex cr. x H1-'19 &  &  & -1.214  &   &  & 0.000  \\ 
			&  &  & (1.528) &   &  & (0.062) \\ 
			sex cr. x H2-'19 &  &  & 0.273   &  &  & 0.003  \\ 
			&  &  & (1.144)  &  &  & (0.117) \\ 
			sex cr. x H1-'20 &  &  & -1.155   &  &  & -0.107  \\ 
			&  &  & (1.407)  &  &  & (0.098) \\ 
			sex cr. x H2-'20 &  &  & -0.621    &  &  & -0.069  \\ 
			&  &  & (1.587)  &  &  & (0.098) \\ 
			post & X &  &   & X &  &  \\ 
			year FE &  & X &   &  & X &  \\ 
			half-year FE &  &  & X  &  &  & X \\ 
			court FE & X & X & X & X & X & X\\ 
			\# words & X & X & X & X & X & X \\ 
			\bottomrule
	\end{tabular}}
	\caption[DiD estimates of effect heterogeneity for victim blaming indicators]{DiD estimates of effect heterogeneity for victim blaming indicators. Significance levels: * p<0.1, ** p<0.05, *** p<0.01.}\label{tab:victimblaming_metoo}
\end{table}

While assessing the impact of the \#MeToo movement on the victim blaming indicators does not yield significant results, the plot on the development of the semantics estimator is nevertheless interesting. It shows that the use of the victim as grammatical subject is generally substantially higher in opinions on sexual assault cases than in cases on other crimes against persons. Given that several studies indicate higher prevalence of victim blaming in sexual assault cases than in cases on other crimes against persons (see Section \ref{victimblaming_metoo}) and that the indicator is constructed based on scientific findings (see Section \ref{victimblamingindicators_metoo}), it seems reasonable to explore whether this indicator may be useful for measuring the extent of victim blaming in judicial opinions - as long as the purpose is to classify opinions or quantify the status quo rather than measure changes over time. The sentiment indicator, on the other hand, seems to vary little both over time and crime types. It generally takes very low values, which may be because judges deliberately avoid sentiment-charged language. It therefore does not seem appropriate for use in the context of court opinions and may be better suited for contexts with more colloquial language such as social media.

The results of the robustness checks  for the DiD with victim blaming indicators are provided in Appendix \ref{app:robustness_metoo}. Given that no significant effects on either indicator can be identified using the DiD approach, the fact that the placebo test does not indicate a violation of the parallel trend assumption is not particularly informative. For the sentiment indicator, both the IPW estimates as well as the DiD estimates based on sodomy and rape cases only indicate a significant decrease in negatively connoted words in the context of mentions of the perpetrator. In the case of the semantics indicator, both estimates have different signs and are not statistically significant.

The effect heterogeneity estimates obtained using the DiD and event study approach, respectively,  can be found in Appendices \ref{app:didheterogeneity_metoo} and \ref{app:EventStudyheterogeneity_metoo}. For the semantics indicator, the estimates suggest a larger decline in mentions of the victim as subject among female and Democratic judges as well as in predominately Democratic states, although again the differences are not statistically significant. The DiD estimates for the sentiment indicator imply a greater decline in the use of words with negative connotations among females and no difference in the effect of \#MeToo on this indicator between Democrats and Republicans. The event study approach, on the other hand, suggests that Democrats have increased their use of negatively connoted words when compared to the development of this indicator among Republicans.

\begin{figure}
	\centering
	\captionsetup{font=footnotesize}
	\includegraphics[width=0.49\linewidth]{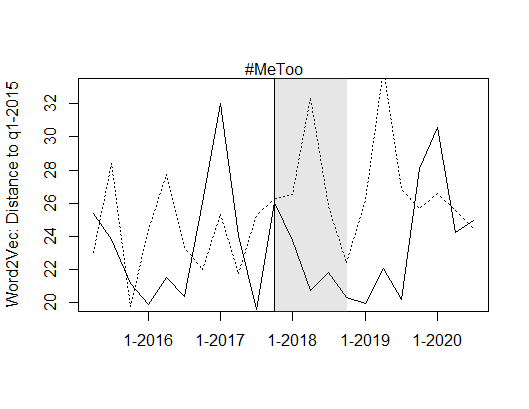}
	\caption[Development of the word2vec representation of the word ``victim'']{\label{fig:word2vec_metoo} Development of the word2vec representation of the word ``victim'', with the solid line representing sexual offense opinions and the dotted line representing the control group.}
\end{figure}

Finally, Figure \ref{fig:word2vec_metoo} displays the development of the Word2Vec representation of the word ``victim'' in both the control and treatment group. Contrary to the research hypothesis, the context in which victim is mentioned does not evolve faster for opinions on sexual offenses than for those on other crimes against persons. Again, this approach may be better suited to contexts with more flexible and rapidly evolving language.

\FloatBarrier

\subsection{Text Vectorization} 
\captionsetup{font=footnotesize}
\begin{figure}
	\centering
	\begin{subfigure}{.49\textwidth}
		\centering
		\includegraphics[width=\linewidth]{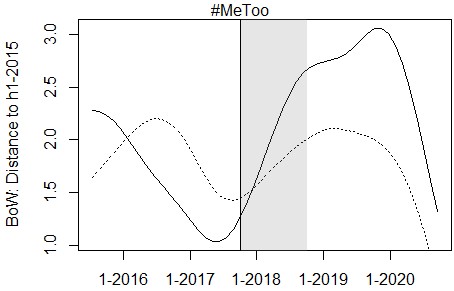}
		\label{fig:sub1_metoo}
	\end{subfigure}%
	\begin{subfigure}{.49\textwidth}
		\centering
		\includegraphics[width=\linewidth]{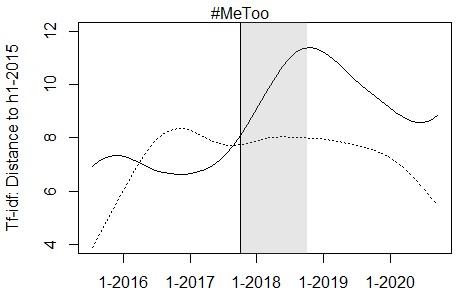}
		\label{fig:sub2_metoo}
	\end{subfigure}
	\caption[Distance of text vectors from their H1 2015 average over time]{Kernel-smoothed plot of the distance of text vectors from their H1 2015 average, with the solid line representing sexual offense opinions and the dotted line representing the control group. The distance of opinions to the H1 2015 average is expressed relative to the median distance of all opinions to the H1 2015 average, in percent.}
	\label{fig:smoothedTextVectors_metoo}
\end{figure}
Figure \ref{fig:smoothedTextVectors_metoo} illustrates the evolution of the BoW\footnote{The reduced sample BoW quantifier evolves similarly to the BoW quantifier in both groups, which is why it is not shown here.} and the tf-idf text quantifiers in opinions on sexual offenses and on other crimes against persons. Both charts suggest that language in opinions on sexual offenses evolves more rapidly than that in opinions on other crimes against persons between the onset of the movement and 2019. However, the graphs also indicate that there may be problems with the parallel trend assumption. Moreover, the BoW graph shows a narrowing of the distance between the opinion vectors and their H1-2015 average in 2016 and 2017, as well as in 2020, which could be due to changes in the composition of courts, but also to other unobservable factors, which in turn would be critical for identifying the causal effect \footnote{The use of other smoothing methods and smaller bandwidths yields similar curves. Thus, it does not seem to be an (over-)smoothing issue.}.

The DiD estimates for the text vectorization-based opinion quantifiers can be found in Table \ref{tab:DiDtextquantifiers_metoo}. The results point at a slight \#MeToo-induced change in courtroom language, which however materializes with a substantial time lag. The estimates suggest that the language in sexual offense opinions deviates more quickly from the 2015 average than in opinions on other cases of crimes against persons. The DiD estimate for the second half of 2020 is statistically insignificant for all three text quantifiers and numerically small for the two BoW quantifiers, indicating that the language change in sexual offense opinions is not likely due to the COVID-19 crisis.

\begin{figure}		
	\captionsetup{font=footnotesize}
	\centering
	\begin{subfigure}{.45\textwidth}
		\centering
		\includegraphics[width=\linewidth]{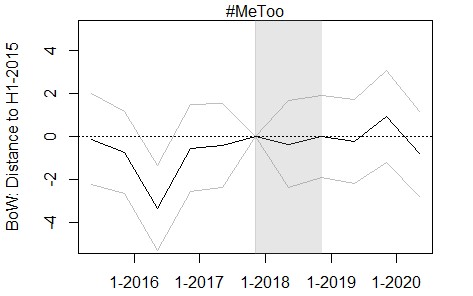}
	\end{subfigure}
	\begin{subfigure}{.45\textwidth}
		\centering
		\includegraphics[width=\linewidth]{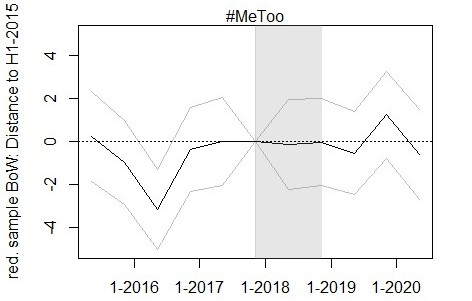}
	\end{subfigure}
	\begin{subfigure}{.45\textwidth}
		\centering
		\includegraphics[width=\linewidth]{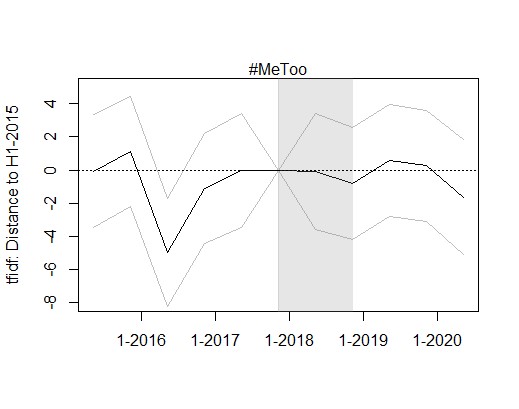}
	\end{subfigure}
	\caption[Event study estimates for evolution of text vectorization-based opinion quantifiers]{\label{fig:event_textquantifiers_metoo} Event study estimates for the  evolution of the text vectorization-based opinion quantifiers. The gray lines indicate the 90 percent confidence interval around the estimates. The distance of opinions to the H1 2015 average is expressed relative to the median distance of all opinions to the H1 2015 average, in percent.}\end{figure}

The event study estimates in Figure \ref{fig:event_textquantifiers_metoo} show a decrease in the distance to the H1-2015 average in 2016 that is similar to, though less pronounced than, that observed in Figure \ref{fig:smoothedTextVectors_metoo}. Further, the event study estimates do not indicate a stronger deviation of language from the H1-2015 average in the years after \#MeToo than in the years before \#MeToo, suggesting that the effect estimated with the DiD approach may be attributable to changes in case composition or personnel rather than changes in judges' attitudes.

\begin{table}
	\captionsetup{font=footnotesize}
	\centering
	\adjustbox{width = \textwidth, center}{%
		\begin{tabular}{lccccccccc}
			
			\toprule
			& \multicolumn{3}{c}{BoW} &
			\multicolumn{3}{c}{Reduced Sample BoW}            &               & tf-idf &           \\ 
			\midrule
			sex cr. x post&0.815*&&&0.700&&&1.665&&\\
			& (0.469) &  &  & (0.511) &  &  & (1.562) &  &  \\ 
			sex cr. x '19 &  & 0.629  &  &  & 0.269  &  &  & 1.338  &  \\ 
			&  & (0.535) &  &  & (0.507) &  &  & (1.732) &  \\ 
			sex cr. x '20 &  & 0.919** &  &  & 1.165** &  &  & 1.106* &  \\ 
			&  & (0.454) &  &  & (0.496) &  &  & (0.646) &  \\ 
			sex cr. x H1-'19 &  &  & 0.658  &  &  & 0.474  &  &  & 0.940  \\ 
			&  &  & (0.715) &  &  & (0.718) &  &  & (1.835) \\ 
			sex cr. x H2-'19 &  &  & 0.603  &  &  & 0.079  &  &  & 1.716  \\ 
			&  &  & (0.563) &  &  & (0.564) &  &  & (1.813) \\ 
			sex cr. x H1-'20 &  &  & 1.461  &  &  & 1.806* &  &  & 0.281  \\ 
			&  &  & (0.905) &  &  & (0.955) &  &  & (0.966) \\ 
			sex cr. x H2-'20 &  &  & 0.317  &  &  & 0.459  &  &  & 1.958  \\ 
			&  &  & (0.692) &  &  & (0.670) &  &  & (1.312) \\ 
			post & X &  &  & X &  &  & X &  &  \\ 
			year FE &  & X &  &  & X &  &  & X &  \\ 
			half-year FE &  &  & X &  &  & X &  &  & X \\ 
			court FE & X & X & X & X & X & X & X & X & X \\ 
			\# words & X & X & X & X & X & X & X & X & X \\ 		\bottomrule
	\end{tabular}}\caption[DiD estimates for text vectorization-based opinion quantifier]{DiD estimates for text vectorization-based opinion quantifiers. The distance of opinions to the H1 2015 average is expressed relative to the median distance of all opinions to the H1 2015 average, in percent. Significance levels: * p<0.1, ** p<0.05, *** p<0.01.}\label{tab:DiDtextquantifiers_metoo}
\end{table}

The results of the robustness checks are presented in Appendix \ref{app:robustness_metoo}. While the placebo tests do not reveal a significant violation of the common trend assumption, they do not allow me to rule out such a violation, especially because the estimated effects of the placebo treatment on the (reduced sample) BoW quantifiers are positive, just like the observed effect in the main DiD analysis. I therefore also estimate the effect of placebo treatments at other points in time during the pre-\#MeToo period, all of which turned out to be statistically insignificant, with some of them having a positive and others a negative sign.  The fact that the IPW-based DiD and the reduced sample DiD estimate positive effects for all quantifiers, some of which are statistically significant, support the finding that the \#MeToo movement has caused a slight increase in language development in sexual assault opinions, with the IPW results ruling out that the observed effect in the main analysis is due to changes in the composition of courts, while the latter rules out that it is due to an increase of reports of sexual offenses. When considering these results in conjunction with the event study results, one possible explanation for the observed slight linguistic change in sexual offense opinions is the change in judicial appointments toward judges with more progressive views on sex offenses. However, given the numerically small effect estimates, the noise in these quantifiers, and the fact that the effect does not appear until two years after the movement began, it is difficult to attribute the observed effect to the \#MeToo movement. 

A look at the effect heterogeneity estimates in Appendix \ref{app:EventStudyheterogeneity_metoo} and \ref{app:didheterogeneity_metoo} does not give a clear picture. While the DiD estimates for the (reduced sample) BoW suggest a faster evolution of language in opinions written by females and Democrats, the event study plots point to a similar evolution of language in opinions written by female and male or Democratic and Republican judges, respectively. For the tf-idf approach, the DiD estimates suggest that the language of female and Democratic judges evolves less rapidly than that of their counterparts.

\section{Conclusion}\label{conclusion_metoo}
In this study, I quantified judicial opinions by means of different indicators and text vectorization methods to assess how the \#MeToo movement has affected the evolution of language in sexual assault opinions and whether it has led to a decrease in victim blaming in such cases. Although I did not obtain statistically significant estimates for the impact of the movement on most quantifiers, with a few exceptions of mildly statistically significant estimates, the point estimates suggest a faster evolution of language in sexual assault opinions as well as a decline in victim blaming.  The reasons for not obtaining statistically significant results may be manifold. For one, the language in judicial opinions is generally not very flexible, i.e., there is regulation on what different parties are allowed to say in court and the structure of court opinions, particularly in certain paragraphs, is highly formalized. Therefore, any treatment should be expected to have a smaller effect on language in court opinions than on language in any other more flexible text body. Moreover, the text vectorization methods in particular, but also the victim blaming indicators, capture a lot of noise, leading to large standard errors in the estimators.

The study's real merit lies in its potential to make a valuable contribution to the expanding body of literature on text-based causal inference: for one, I have developed indicators that can be useful proxies for victim blaming and may be used as a treatment, outcome or control, if they are applied to a text body with more flexible language. Then, I have also introduced an approach to quantifying language development that is based on text vectorization methods originally designed for categorization and clustering purposes. This enables the use of text vectorization methods in the context of DiD analyses or panel data methods. Again, this approach may be useful for assessing text bodies with more flexible language, analyzing larger text bodies, or evaluating longer-term effects of a treatment in a panel setting. Moreover, the victim-blaming indicators may be used in descriptive contexts to gauge the level of victim blaming in a text body or to compare the extent of victim blaming across different text bodies.

\pagebreak
\bibliographystyle{econometrica}
\bibliography{Literatura}

\begin{appendix}
	\section{Descriptives}
	\begin{table}[H]
		\captionsetup{font=footnotesize}
		\centering
		\adjustbox{width = 1.05\textwidth, center}{%
			\begin{tabular}{lcc}
				\toprule
				Court & \# Non-sexual Offenses & \# Sexual Offenses \\ 
				\midrule
				Appellate Court of Illinois & 668 & 192 \\ 
				Army Court of Criminal Appeals &  10 &  46 \\ 
				California Court of Appeal & 912 & 361 \\ 
				Commonwealth Court of Pennsylvania & 247 & 114 \\ 
				Connecticut Appellate Court & 436 & 164 \\ 
				Court of Appeals for the D.C. Circuit &  60 &  24 \\ 
				Court of Appeals for the Eighth Circuit & 349 & 194 \\ 
				Court of Appeals for the Eleventh Circuit & 179 &  64 \\ 
				Court of Appeals for the Federal Circuit &  14 &   3 \\ 
				Court of Appeals for the Fifth Circuit & 259 & 108 \\ 
				Court of Appeals for the First Circuit & 147 &  70 \\ 
				Court of Appeals for the Fourth Circuit & 167 &  61 \\ 
				Court of Appeals for the Ninth Circuit & 242 &  98 \\ 
				Court of Appeals for the Second Circuit & 115 &  44 \\ 
				Court of Appeals for the Seventh Circuit & 367 & 134 \\ 
				Court of Appeals for the Sixth Circuit & 188 &  95 \\ 
				Court of Appeals for the Tenth Circuit & 137 &  51 \\ 
				Court of Appeals for the Third Circuit &  98 &  36 \\ 
				Court of Appeals of Alaska &  47 &  31 \\ 
				Court of Appeals of Arizona &  43 &  31 \\ 
				Court of Appeals of Arkansas & 192 & 156 \\ 
				Court of Appeals of Georgia & 403 & 375 \\ 
				Court of Appeals of Iowa & 1244 & 680 \\ 
				Court of Appeals of Kansas &  48 &  47 \\ 
				Court of Appeals of Minnesota &   1 &  19 \\ 
				Court of Appeals of Mississippi & 426 & 197 \\ 
				\bottomrule
			\end{tabular}
			
			\begin{tabular}{lcc}
				\toprule
				Court & \# Non-sexual Offenses & \# Sexual Offenses \\ 
				\midrule
				Court of Appeals of North Carolina & 274 & 147 \\ 
				Court of Appeals of Tennessee & 121 &  85 \\ 
				Court of Appeals of Texas & 4082 & 2576 \\ 
				Court of Appeals of Virginia &  92 &  39 \\ 
				Court of Appeals of Washington & 204 & 137 \\ 
				Court of Criminal Appeals of Tennessee & 1949 & 797 \\ 
				Court of Criminal Appeals of Texas & 252 & 120 \\ 
				District Court of Appeal of Florida & 1123 & 324 \\ 
				District Court, District of Columbia & 366 & 133 \\ 
				District of Columbia Court of Appeals & 154 &  34 \\ 
				Idaho Court of Appeals &  48 &  42 \\ 
				Indiana Court of Appeals & 1781 & 1146 \\ 
				Massachusetts Appeals Court &  45 &  98 \\ 
				Michigan Court of Appeals &  85 &  67 \\ 
				Missouri Court of Appeals & 354 & 285 \\ 
				Navy-Marine Corps Court of Criminal Appeals &  76 & 252 \\ 
				Nebraska Court of Appeals &  99 & 113 \\ 
				New Jersey Superior Court &  65 &  35 \\ 
				New Mexico Court of Appeals &  60 &  33 \\ 
				New York Court of Appeals &  93 &  40 \\ 
				Ohio Court of Appeals & 3283 & 2091 \\ 
				Superior Court of Delaware & 159 &  36 \\ 
				Superior Court of Pennsylvania & 5995 & 3246 \\ 
				United States Air Force Court of Criminal Appeals &   3 &  28 \\ 
				United States Court of Federal Claims &  19 &   8 \\ 
				&& \\
				\bottomrule
		\end{tabular}}\caption{Number of opinions per court.}\label{tab:descCourt_metoo}
	\end{table}
	\FloatBarrier

	\section{DiD: Robustness Checks}\label{app:robustness_metoo}
	\subsection{Victim Blaming Indicators}
	\begin{table}[H]
		\captionsetup{font=footnotesize}
		\begin{singlespace}
			\centering
			\adjustbox{width = 0.9\textwidth, center}{%
				
				\begin{tabular}{lcccccc}
					\toprule
					& \multicolumn{3}{c}{Victim as Subject} &               \multicolumn{3}{c}{Neg. of Offender Context}       \\ 
					& (1) & (2) & (3) & (1) & (2) & (3) \\
					\midrule
					sex cr. x placebo & -0.716  &  &  & -0.012  &  &  \\ 
					& (0.810) &  &  & (0.066) &  &  \\ 
					sex cr. x post &  & -0.939  &  &  & -0.100* &  \\ 
					&  & (0.926) &  &  & (0.058) &  \\ 
					sex cr. x post &  &  & 1.399  &  &  & -0.093** \\ 
					&  &  & (1.564) &  &  & (0.042) \\ 
					post & X & X & X & X & X & X \\ 
					court FE & X & X & X & X & X & X \\ 
					\# words & X & X & X & X & X & X \\ 
					\bottomrule
			\end{tabular}}
			\caption[DiD robustness tests, victim blaming indicators]{Robustness tests: (1) DiD was performed using only pretreatment observations and a placebo treatment in the middle of the pretreatment period; (2) DiD with IPW based on court distribution, performed for the entire sample using the \texttt{didweight} function from the \texttt{causalweight} package in the statistical software R (\cite*{R2022}); and (3) DiD performed only for the sample of sodomy and sexual assault cases. Significance levels: * p<0.1, ** p<0.05, *** p<0.01.}
		\end{singlespace}
	\end{table}
	
	\FloatBarrier
	\subsection{Text Vectorization}
	\begin{table}[ht]
		\captionsetup{font=footnotesize}
		\begin{singlespace}
			\centering
			\adjustbox{width = \textwidth, center}{%
				\begin{tabular}{lccccccccc}
					\toprule
					& \multicolumn{3}{c} {BoW} &
					\multicolumn{3}{c}{Reduced Sample BoW}            &               \multicolumn{3}{c}{tf-idf}       \\ 
					& (1) & (2) & (3) & (1) & (2) & (3) & (1) & (2) & (3) \\
					\midrule
					sex cr. x placebo & 0.220  &  &  & 0.166  &  &  & -0.941  &  &  \\ 
					& (0.584) &  &  & (0.675) &  &  & (0.831) &  &  \\ 
					sex cr. x post &  & 0.535  &  &  & 0.442  &  &  & 1.697** &  \\ 
					&  & (0.424) &  &  & (0.438) &  &  & (0.821) &  \\ 
					sex cr. x post &  &  & 0.806  &  &  & 0.836* &  &  & 4.148* \\ 
					&  &  & (0.534) &  &  & (0.505) &  &  & (2.254) \\ 
					post & X & X & X & X & X & X & X & X & X \\ 
					court FE & X & X & X & X & X & X & X & X & X \\ 
					\# words & X & X & X & X & X & X & X & X & X \\ 
					\bottomrule
			\end{tabular}}
			\caption[DiD robustness tests, text vectorization-based opinions quantifiers]{Robustness tests: (1) DiD was performed using only pretreatment observations and a placebo treatment in the middle of the pretreatment period; (2) DiD with IPW based on court distribution, performed for the entire sample using the \texttt{didweight} function from the \texttt{causalweight} package in R; and (3) DiD performed only for the sample of sodomy and sexual assault cases. The distance of opinions to the H1 2015 average is expressed relative to the median distance of all opinions to the H1 2015 average, in percent. Significance levels: * p<0.1, ** p<0.05, *** p<0.01. }
		\end{singlespace}
	\end{table}

	\FloatBarrier
	\pagebreak
	\section{DiD: Effect Heterogeneity}\label{app:didheterogeneity_metoo}
	
	\subsection{Victim Blaming Indicators}
	\begin{table}[H]
		\captionsetup{font=footnotesize}
		\begin{singlespace}
			\centering
			\adjustbox{width = 0.9\textwidth, center}{%
				\begin{tabular}{lcccccc}
					\toprule
					& \multicolumn{3}{c}{Victim as Subject} &               \multicolumn{3}{c}{Neg. of Offender Context}       \\ 
					\midrule
					female x sexual x post & -1.103  &  &  & -0.236  &  &  \\ 
					& (3.329) &  &  & (0.394) &  &  \\ 
					dem. judge x sexual x post &  & -2.099  &  &  & -0.015  &  \\ 
					&  & (3.729) &  &  & (0.225) &  \\ 
					dem. state x sexual x post &  &  & -0.276  &  &  & -0.106  \\ 
					&  &  & (2.967) &  &  & (0.162) \\ 
					post & X & X & X & X & X & X \\ 
					court FE & X & X & X & X & X & X \\ 
					\# words & X & X & X & X & X & X \\ 
					\bottomrule
			\end{tabular}} \caption[DiD estimates of effect heterogeneity, victim blaming indicators]{DiD estimates of effect heterogeneity for victim blaming indicators. Significance levels: * p<0.1, ** p<0.05, *** p<0.01.}
		\end{singlespace}
	\end{table}
	
	\FloatBarrier
	\subsection{Text Vectorization}
	\begin{table}[H]
		\captionsetup{font=footnotesize}
		\begin{singlespace}
			\centering
			\adjustbox{width = 1.05\textwidth, center}{%
				\begin{tabular}{lccccccccc}
					\toprule
					& \multicolumn{3}{c} {BoW} &
					\multicolumn{3}{c}{Reduced Sample BoW}            &               \multicolumn{3}{c}{tf-idf}        \\ 
					\midrule
					female x sexual x post & 0.891  &  &  & 1.170  &  &  & -1.740  &  &  \\ 
					& (1.171) &  &  & (1.194) &  &  & (2.207) &  &  \\ 
					dem. judge x sexual x post &  & 0.340  &  &  & 0.764  &  &  & -4.144  &  \\ 
					&  & (1.938) &  &  & (2.145) &  &  & (2.993) &  \\ 
					dem. state x sexual x post &  &  & 0.357  &  &  & -0.152  &  &  & 2.169  \\ 
					&  &  & (1.335) &  &  & (1.207) &  &  & (1.841) \\ 
					post & X & X & X & X & X & X & X & X & X \\ 
					court FE & X & X & X & X & X & X & X & X & X \\ 
					\# words & X & X & X & X & X & X & X & X & X \\ 
					\bottomrule
			\end{tabular}}\caption[DiD estimates of effect heterogeneity, text vectorization-based opinion quantifiers]{DiD estimates of effect heterogeneity for text vectorization-based opinion quantifiers. The distance of opinions to the H1 2015 average is expressed relative to the median distance of all opinions to the H1 2015 average, in percent. Significance levels: * p<0.1, ** p<0.05, *** p<0.01.}
		\end{singlespace}
	\end{table}
	\FloatBarrier
	
	\pagebreak
	\section{Event Study Approach: Effect Heterogeneity}\label{app:EventStudyheterogeneity_metoo}
	
	\subsection{By Gender}
	\begin{figure}[ht]
		\captionsetup{font=footnotesize}
		\centering
		\begin{subfigure}{.45\textwidth}
			\centering
			\includegraphics[width=\linewidth]{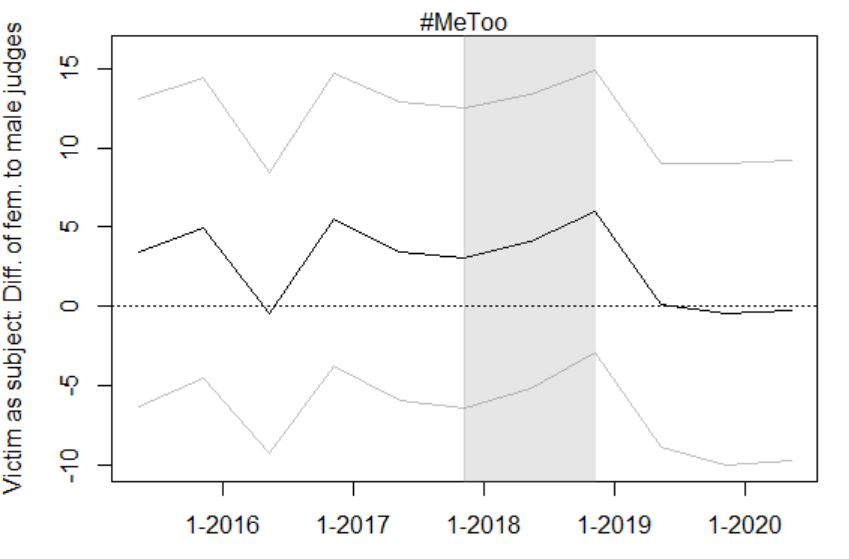}
		\end{subfigure}
		\begin{subfigure}{.47\textwidth}
			\centering
			\includegraphics[width=\linewidth]{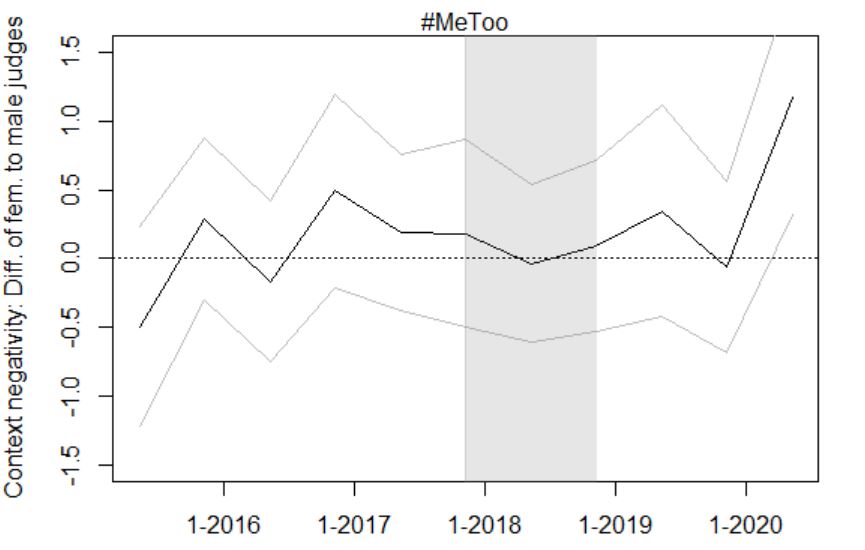}
		\end{subfigure}
		
		\begin{subfigure}{.45\textwidth}
			\centering
			\includegraphics[width=\linewidth]{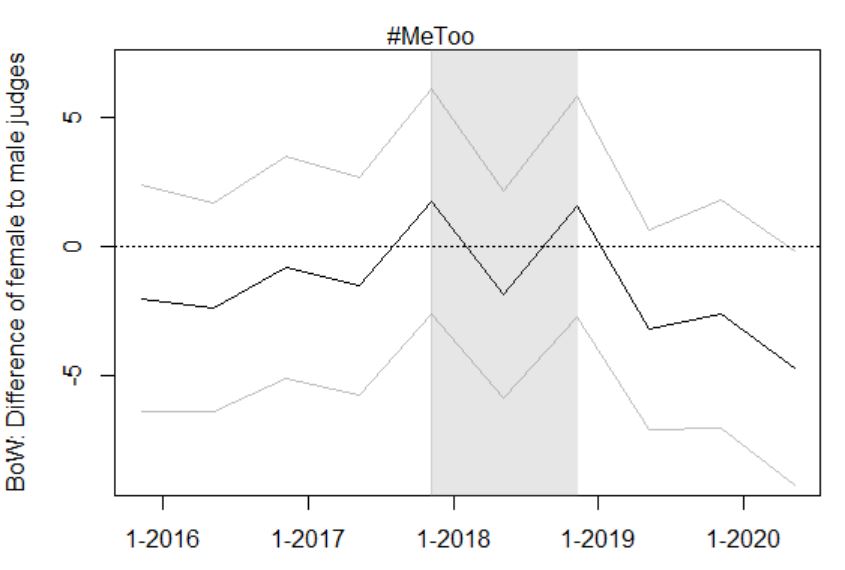}
		\end{subfigure}
		\begin{subfigure}{.45\textwidth}
			\centering
			\includegraphics[width=\linewidth]{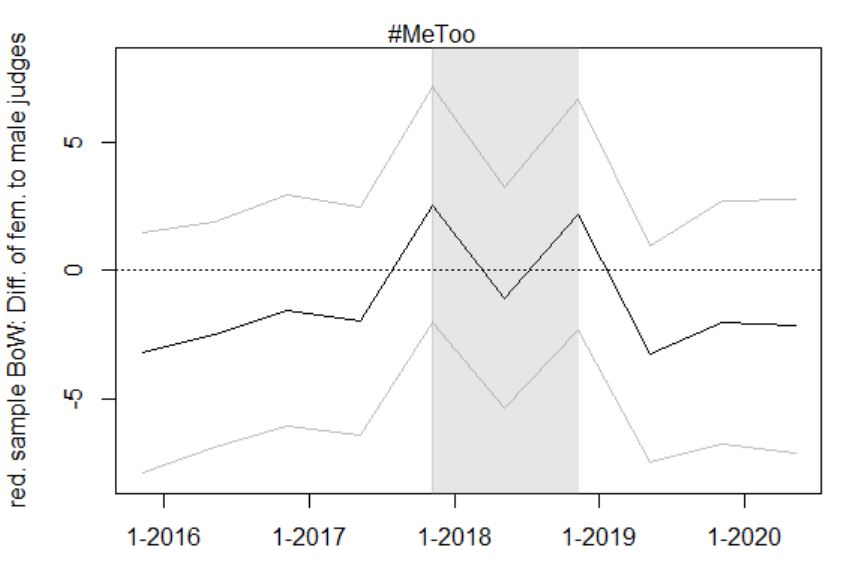}
		\end{subfigure}
		\begin{subfigure}{.45\textwidth}
			\centering
			\includegraphics[width=\linewidth]{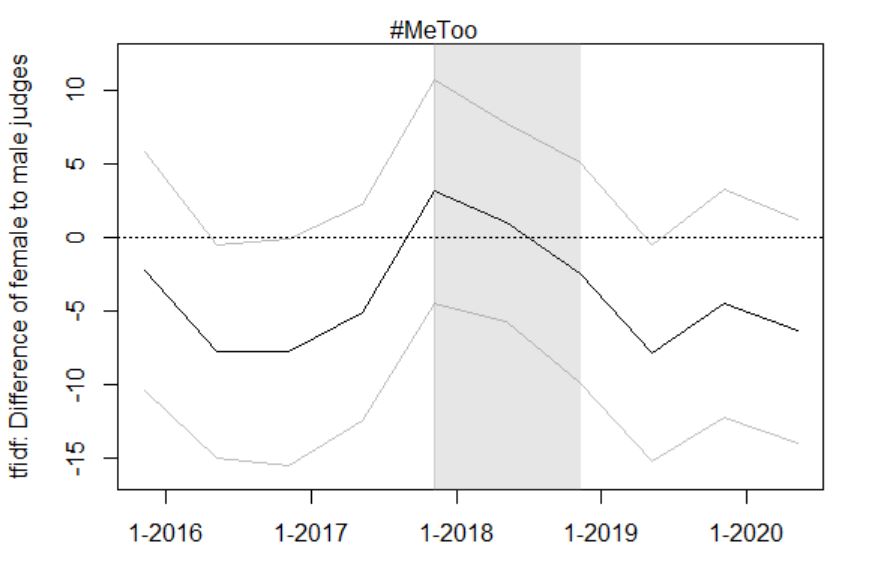}
		\end{subfigure}
		\caption[Event study estimates of difference in development of female vs. male judges]{Event study estimates of the difference in indicator development of female vs. male judges. The black line represents the FE estimates for the interaction terms of 6-month period identifiers and a dummy variable indicating whether a judge is female. The gray lines indicate the 90 percent confidence interval around the estimates. }
	\end{figure}
	\pagebreak
	\FloatBarrier
	
	\subsection{By Political Affiliation}
	\begin{figure}[ht]
		\captionsetup{font=footnotesize}
		\centering
		\begin{subfigure}{.45\textwidth}
			\centering
			\includegraphics[width=\linewidth]{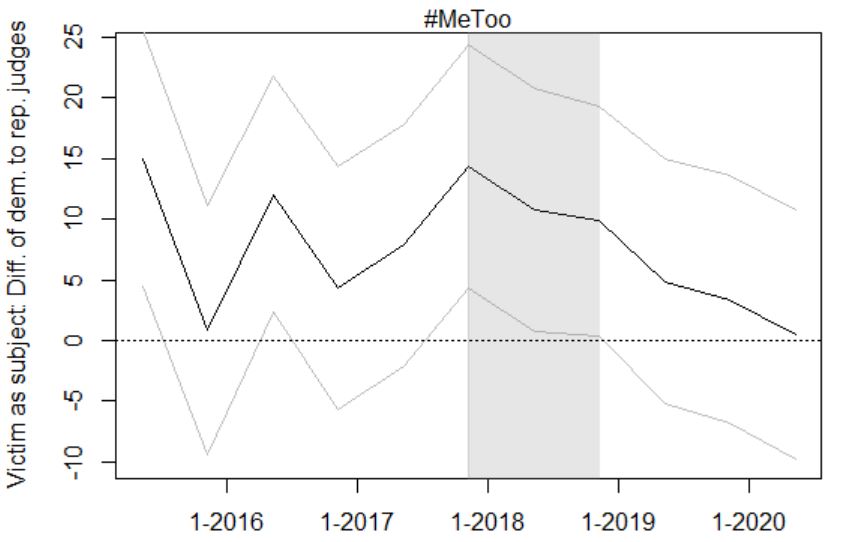}
		\end{subfigure}
		\begin{subfigure}{.45\textwidth}
			\centering
			\includegraphics[width=\linewidth]{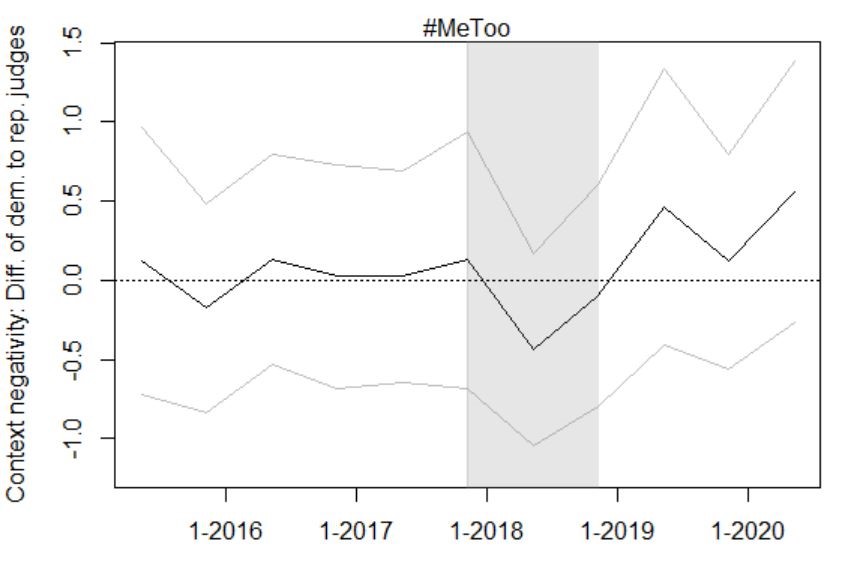}
		\end{subfigure}
		
		\begin{subfigure}{.45\textwidth}
			\centering
			\includegraphics[width=\linewidth]{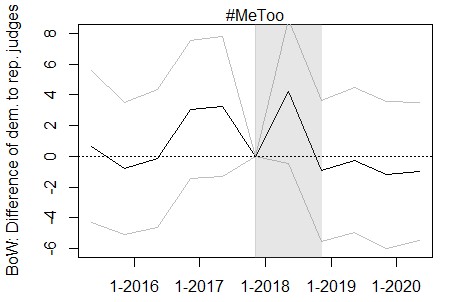}
		\end{subfigure}
		\begin{subfigure}{.45\textwidth}
			\centering
			\includegraphics[width=\linewidth]{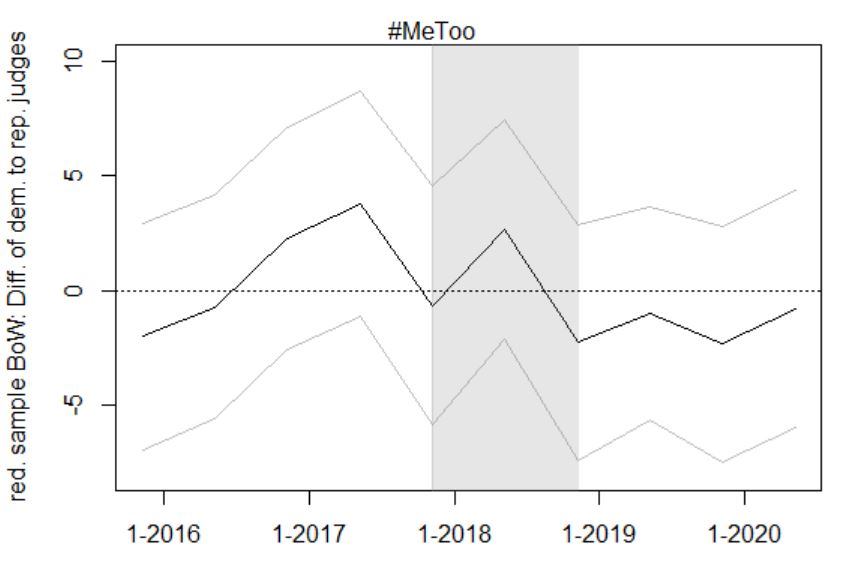}
		\end{subfigure}
		\begin{subfigure}{.45\textwidth}
			\centering
			\includegraphics[width=\linewidth]{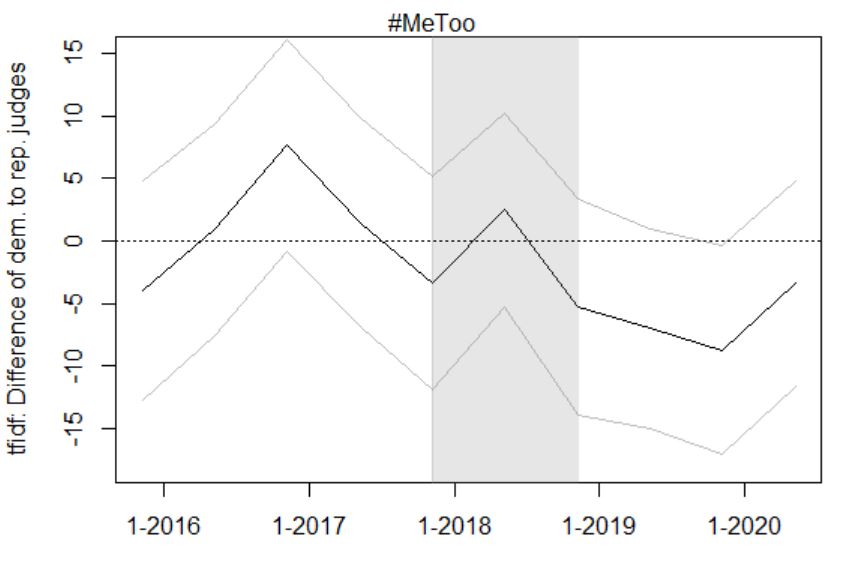}
		\end{subfigure}
		\caption[Event study estimates of difference in development of Democrats vs. Republicans]{Event study estimates of the difference in indicator development of Democratic vs. Republican judges. The black line represents the FE estimates for the interaction terms of 6-month period identifiers and a dummy variable indicating whether a judge is affiliated with the Democratic party. The gray lines indicate the 90 percent confidence interval around the estimates.}
	\end{figure}
	
\end{appendix}

\end{document}